\documentclass[a4paper,12pt]{article}
\usepackage{graphicx}
\begin{document}

\newcommand{\be}{\begin{equation}}
\newcommand{\beq}{\begin{equation}}
\newcommand{\eeq}{\end{equation}}
\newcommand{\ee}{\end{equation}}

\newcommand{\beqn}{\begin{eqnarray}}
\newcommand{\eeqn}{\end{eqnarray}}
\newcommand{\bea}{\begin{eqnarray}}
\newcommand{\ena}{\end{eqnarray}}
\newcommand{\ra}{\rightarrow}
\newcommand{\susy}{{{\cal SUSY}$\;$}}
\newcommand{\su}{$ SU(2) \times U(1)\,$}

\newcommand{\gag}{$\gamma \gamma$ }
\newcommand{\gagt}{\gamma \gamma }
\newcommand{\gam}{\gamma \gamma }
\def\W{{\mbox{\boldmath $W$}}}
\def\B{{\mbox{\boldmath $B$}}}
\def\V{{\mbox{\boldmath $V$}}}
\newcommand{\np}{Nucl.\,Phys.\,}
\newcommand{\pl}{Phys.\,Lett.\,}
\newcommand{\pr}{Phys.\,Rev.\,}
\newcommand{\prl}{Phys.\,Rev.\,Lett.\,}
\newcommand{\prep}{Phys.\,Rep.\,}
\newcommand{\zp}{Z.\,Phys.\,}
\newcommand{\sovjnp}{{\em Sov.\ J.\ Nucl.\ Phys.\ }}
\newcommand{\nuclinst}{{\em Nucl.\ Instrum.\ Meth.\ }}
\newcommand{\annp}{{\em Ann.\ Phys.\ }}
\newcommand{\intjmp}{{\em Int.\ J.\ of Mod.\  Phys.\ }}

\newcommand{\eps}{\epsilon}
\newcommand{\mw}{M_{W}}
\newcommand{\mww}{M_{W}^{2}}
\newcommand{\mwmw}{M_{W}^{2}}
\newcommand{\mhmh}{M_{H}^2}
\newcommand{\mz}{M_{Z}}
\newcommand{\mzz}{M_{Z}^{2}}

\newcommand{\cw}{\cos\theta_W}
\newcommand{\sw}{\sin\theta_W}
\newcommand{\tw}{\tan\theta_W}
\def\cww{\cos^2\theta_W}
\def\sww{\sin^2\theta_W}
\def\tww{\tan^2\theta_W}

\newcommand{\epm}{$e^{+} e^{-}\;$}
\newcommand{\epemt}{$e^{+} e^{-}\;$}
\newcommand{\epem}{e^{+} e^{-}\;}
\newcommand{\ememt}{$e^{-} e^{-}\;$}
\newcommand{\emem}{e^{-} e^{-}\;}

\newcommand{\lra}{\leftrightarrow}
\newcommand{\tr}{{\rm Tr}}
\def\ls1{{\not l}_1}
\newcommand{\cms}{centre-of-mass\hspace*{.1cm}}


\newcommand{\dkg}{\Delta \kappa_{\gamma}}
\newcommand{\dkz}{\Delta \kappa_{Z}}
\newcommand{\dz}{\delta_{Z}}
\newcommand{\dgz}{\Delta g^{1}_{Z}}
\newcommand{\dgzt}{$\Delta g^{1}_{Z}\;$}
\newcommand{\la}{\lambda}
\newcommand{\lag}{\lambda_{\gamma}}
\newcommand{\lambdae}{\lambda_{e}}
\newcommand{\laz}{\lambda_{Z}}
\newcommand{\lnl}{L_{9L}}
\newcommand{\lnr}{L_{9R}}
\newcommand{\lt}{L_{10}}
\newcommand{\lu}{L_{1}}
\newcommand{\ld}{L_{2}}
\newcommand{\eeww}{e^{+} e^{-} \ra W^+ W^- \;}
\newcommand{\eewwt}{$e^{+} e^{-} \ra W^+ W^- \;$}
\newcommand{\epemww}{e^{+} e^{-} \ra W^+ W^- }
\newcommand{\epemwwt}{$e^{+} e^{-} \ra W^+ W^- \;$}
\newcommand{\eennhht}{$e^{+} e^{-} \ra \nu_e \bar \nu_e HH\;$}
\newcommand{\eennhh}{e^{+} e^{-} \ra \nu_e \bar \nu_e HH\;}
\newcommand{\ppwg}{p p \ra W \gamma}
\newcommand{\wwhh}{W^+ W^- \ra HH\;}
\newcommand{\wwhht}{$W^+ W^- \ra HH\;$}
\newcommand{\ppwz}{pp \ra W Z}
\newcommand{\ppwgt}{$p p \ra W \gamma \;$}
\newcommand{\ppwzt}{$pp \ra W Z \;$}
\newcommand{\gamgamt}{$\gamma \gamma \;$}
\newcommand{\gamgam}{\gamma \gamma \;}
\newcommand{\egamt}{$e \gamma \;$}
\newcommand{\egam}{e \gamma \;}
\newcommand{\gamgamwwt}{$\gamma \gamma \ra W^+ W^- \;$}
\newcommand{\gamgamwwht}{$\gamma \gamma \ra W^+ W^- H \;$}
\newcommand{\gamgamwwh}{\gamma \gamma \ra W^+ W^- H \;}
\newcommand{\gamgamwwhht}{$\gamma \gamma \ra W^+ W^- H H\;$}
\newcommand{\gamgamwwhh}{\gamma \gamma \ra W^+ W^- H H\;}
\newcommand{\ggww}{\gamma \gamma \ra W^+ W^-}
\newcommand{\ggwwt}{$\gamma \gamma \ra W^+ W^- \;$}
\newcommand{\ggwwht}{$\gamma \gamma \ra W^+ W^- H \;$}
\newcommand{\ggwwh}{\gamma \gamma \ra W^+ W^- H \;}
\newcommand{\ggwwhht}{$\gamma \gamma \ra W^+ W^- H H\;$}
\newcommand{\ggwwhh}{\gamma \gamma \ra W^+ W^- H H\;}
\newcommand{\ggwwz}{\gamma \gamma \ra W^+ W^- Z\;}
\newcommand{\ggwwzt}{$\gamma \gamma \ra W^+ W^- Z\;$}

\newcommand{\ptu}{p_{1\bot}}
\newcommand{\vecptu}{\vec{p}_{1\bot}}
\newcommand{\ptd}{p_{2\bot}}
\newcommand{\vecptd}{\vec{p}_{2\bot}}
\newcommand{\ie}{{\em i.e.}}
\newcommand{\cm}{{{\cal M}}}
\newcommand{\cl}{{{\cal L}}}
\newcommand{\cd}{{{\cal D}}}
\newcommand{\cv}{{{\cal V}}}
\def\slashc{c\kern -.400em {/}}
\def\slashp{p\kern -.400em {/}}
\def\slashL{L\kern -.450em {/}}
\def\slashcl{\cl\kern -.600em {/}}
\def\Ww{{\mbox{\boldmath $W$}}}
\def\B{{\mbox{\boldmath $B$}}}
\def\noi{\noindent}
\def\nn{\noindent}
\def\sm{${\cal{S}} {\cal{M}}\;$}
\def\smn{${\cal{S}} {\cal{M}}$}
\def\nph{${\cal{N}} {\cal{P}}\;$}
\def\sb{$ {\cal{S}}  {\cal{B}}\;$}
\def\ssb{${\cal{S}} {\cal{S}}  {\cal{B}}\;$}
\def\ssbe{{\cal{S}} {\cal{S}}  {\cal{B}}}
\def\cviol{${\cal{C}}\;$}
\def\pviol{${\cal{P}}\;$}
\def\cpviol{${\cal{C}} {\cal{P}}\;$}

\newcommand{\lgg}{\lambda_1\lambda_2}
\newcommand{\lww}{\lambda_3\lambda_4}
\newcommand{\ppin}{ P^+_{12}}
\newcommand{\pmin}{ P^-_{12}}
\newcommand{\ppout}{ P^+_{34}}
\newcommand{\pmout}{ P^-_{34}}
\newcommand{\sinsq}{\sin^2\theta}
\newcommand{\cossq}{\cos^2\theta}
\newcommand{\yt}{y_\theta}
\newcommand{\hppll}{++;00}
\newcommand{\hpmll}{+-;00}
\newcommand{\hpplt}{++;\lambda_30}
\newcommand{\hpmlt}{+-;\lambda_30}
\newcommand{\hpptt}{++;\lambda_3\lambda_4}
\newcommand{\hpmtt}{+-;\lambda_3\lambda_4}
\newcommand{\dk}{\Delta\kappa}
\newcommand{\klam}{\Delta\kappa \lambda_\gamma }
\newcommand{\kac}{\Delta\kappa^2 }
\newcommand{\lac}{\lambda_\gamma^2 }
\def\gamgamtzz{$\gamma \gamma \ra ZZ \;$}
\def\gamgamtww{$\gamma \gamma \ra W^+ W^-\;$}
\def\gamgamtwwe{\gamma \gamma \ra W^+ W^-}
\def\sinb{\sin\beta}
\def\cosb{\cos\beta}
\def\sinbb{\sin (2\beta)}
\def\cosbb{\cos (2 \beta)}
\def\tgb{\tan \beta}
\def\tgbt{$\tan \beta\;\;$}
\def\tgbsq{\tan^2 \beta}
\def\sinal{\sin\alpha}
\def\cosal{\cos\alpha}
\def\stop{\tilde{t}}
\def\sto{\tilde{t}_1}
\def\stt{\tilde{t}_2}
\def\stl{\tilde{t}_L}
\def\str{\tilde{t}_R}
\def\msto{m_{\sto}}
\def\mstosq{m_{\sto}^2}
\def\mstt{m_{\stt}}
\def\msttsq{m_{\stt}^2}
\def\mt{m_t}
\def\mtsq{m_t^2}
\def\sint{\sin\theta_{\stop}}
\def\sintt{\sin 2\theta_{\stop}}
\def\cost{\cos\theta_{\stop}}
\def\sintsq{\sin^2\theta_{\stop}}
\def\costsq{\cos^2\theta_{\stop}}
\def\mqtt{\M_{\tilde{Q}_3}^2}
\def\mutt{\M_{\tilde{U}_{3R}}^2}
\def\sbottom{\tilde{b}}
\def\sbo{\tilde{b}_1}
\def\sbt{\tilde{b}_2}
\def\sbl{\tilde{b}_L}
\def\sbr{\tilde{b}_R}
\def\msbo{m_{\sbo}}
\def\msbosq{m_{\sbo}^2}
\def\msbt{m_{\sbt}}
\def\msbtsq{m_{\sbt}^2}
\def\mt{m_t}
\def\mtsq{m_t^2}
\def\selectron{\tilde{e}}
\def\seo{\tilde{e}_1}
\def\set{\tilde{e}_2}
\def\sel{\tilde{e}_L}
\def\ser{\tilde{e}_R}
\def\mseo{m_{\seo}}
\def\mseosq{m_{\seo}^2}
\def\mset{m_{\set}}
\def\msetsq{m_{\set}^2}
\def\msel{m_{\sel}}
\def\mser{m_{\ser}}
\def\me{m_e}
\def\mesq{m_e^2}
\def\snu{\tilde{\nu}}
\def\snue{\tilde{\nu_e}}
\def\set{\tilde{e}_2}
\def\snul{\tilde{\nu}_L}
\def\msnue{m_{\snue}}
\def\msnuesq{m_{\snue}^2}
\def\smuon{\tilde{\mu}}
\def\smul{\tilde{\mu}_L}
\def\smur{\tilde{\mu}_R}
\def\msmul{m_{\smul}}
\def\msmulsq{m_{\smul}^2}
\def\msmur{m_{\smur}}
\def\msmursq{m_{\smur}^2}
\def\stau{\tilde{\tau}}
\def\stauo{\tilde{\tau}_1}
\def\staut{\tilde{\tau}_2}
\def\staul{\tilde{\tau}_L}
\def\staur{\tilde{\tau}_R}
\def\mstauo{m_{\stauo}}
\def\mstauosq{m_{\stauo}^2}
\def\mstaut{m_{\staut}}
\def\mstautsq{m_{\staut}^2}
\def\mtau{m_\tau}
\def\mtausq{m_\tau^2}
\def\gluino{\tilde{g}}
\def\mgluino{m_{\tilde{g}}}
\def\mchi{m_\chi^+}
\def\neuto{\tilde{\chi}_1^0}
\def\mneuto{m_{\tilde{\chi}_1^0}}
\def\neutt{\tilde{\chi}_2^0}
\def\mneutt{m_{\tilde{\chi}_2^0}}
\def\neutth{\tilde{\chi}_3^0}
\def\mneutth{m_{\tilde{\chi}_3^0}}
\def\neutf{\tilde{\chi}_4^0}
\def\mneutf{m_{\tilde{\chi}_4^0}}
\def\chargop{\tilde{\chi}_1^+}
\def\mchargo{m_{\tilde{\chi}_1^+}}
\def\chargtp{\tilde{\chi}_2^+}
\def\mchargt{m_{\tilde{\chi}_2^+}}
\def\chargom{\tilde{\chi}_1^-}
\def\chargtm{\tilde{\chi}_2^-}
\def\bino{\tilde{b}}
\def\wino{\tilde{w}}
\def\photino{\tilde{\gamma}}
\def\zino{tilde{z}}
\def\sdowno{\tilde{d}_1}
\def\sdownt{\tilde{d}_2}
\def\sdownl{\tilde{d}_L}
\def\sdownr{\tilde{d}_R}
\def\supo{\tilde{u}_1}
\def\supt{\tilde{u}_2}
\def\supl{\tilde{u}_L}
\def\supr{\tilde{u}_R}
\def\mh{m_h}
\def\mht{m_h^2}
\def\MH{M_H}
\def\MHt{M_H^2}
\def\MA{M_A}
\def\MAt{M_A^2}
\def\MHp{M_H^+}
\def\MHm{M_H^-}

\begin{titlepage}
\def\baselinestretch{1.2}
\vspace*{\fill}
\begin{center}
{\large {\bf {\em SUSY Higgs at the LHC:
\\ Effects of light
charginos and neutralinos.}}}

\vspace*{1.cm}

\begin{tabular}[t]{c}

{\bf G.~B\'elanger$^{1}$, F.~Boudjema$^{1}$, F.~Donato$^{1}$,
R.~Godbole$^{2}$ and S.~Rosier-Lees$^{3}$}
\\
\\
\\
{\it 1. Laboratoire de Physique Th\'eorique} {\large
LAPTH}\footnote{URA 14-36 du CNRS, associ\'ee  \`a l'Universit\'e
de Savoie.}\\
 {\it Chemin de Bellevue, B.P. 110, F-74941 Annecy-le-Vieux,
Cedex, France.}\\

 {\it 2. Centre of Theoretical Studies, Indian
Institute of Science}
\\ {\it Bangalore 560 012, India }\\

{\it 3. Laboratoire de Physique des Particules} {\large
LAPP}\footnote{UMR du CNRS, associ\'ee  \`a l'Universit\'e de
Savoie.}\\
 {\it Chemin de Bellevue, B.P. 110, F-74941 Annecy-le-Vieux,
Cedex, France.}
\end{tabular}
\end{center}

\centerline{ {\bf Abstract} } \baselineskip=14pt \noindent
{\small In view of the latest LEP data we consider the effects of
charginos and neutralinos on the two-photon and $b \bar b$
signatures of the Higgs at the LHC. Assuming the usual GUT
inspired relation between $M_1$ and $M_2$ we show that there are
only small regions with moderate $\tgb$ and  large stop mixings
that may be dangerous. Pathological models not excluded by LEP
which have degeneracy between the sneutrino and the chargino are
however a real danger because of large branching fraction of the
Higgs into invisibles. We have also studied models where the
gaugino masses are not unified at the GUT scale. We take
$M_1=M_2/10$ as an example where large reductions in the signal at
the LHC can occur. However we argue that such models with a very
light neutralino LSP may give a too large relic density unless the
sleptons are light. We then combine this cosmological constraint
with neutralino production with light sfermions to further reduce
the parameter space that precludes observability of the Higgs at
the LHC. We still find regions of parameter space where the drops
in the usual Higgs signals at the LHC can be drastic. Nonetheless,
in such scenarios where Higgs may escape detection we show that
one should be able to produce all charginos and neutralinos.
Although the heavier of these could cascade into the Higgs, the
rates are not too high and the Higgs may not always be recovered
this way.}
\vspace*{\fill}

\vspace*{0.1cm}\rightline{IISc-CTS/3/00}
\vspace*{0.1cm}\rightline{LAPTH-774/2000}

\vspace*{.7cm}
\rightline{{\large January 2000}}

\end{titlepage}
\baselineskip=18pt

\setcounter{section}{0}
 \setcounter{subsection}{0}
\setcounter{equation}{0}
\def\thesubsection {\thesection.\arabic{subsection}}
\def\theequation{\thesection.\arabic{equation}}

\section{Introduction}
Uncovering the mechanism of symmetry breaking is one of the major
tasks of the high energy colliders. Most prominent is the search
for the Higgs particle. Within the standard model, \smn, this
scalar particle poses the problem of naturalness and its mass is a
free parameter. Current data\cite{mh_limit-nov99} seem to indicate
a preference for a light Higgs with a mass that can nicely fit
within a supersymmetric version of the \smn. In fact an
intermediate mass Higgs, IMH, is  one of the most robust
prediction of SUSY, since one does not have strict predictions on
the large array of the other masses and parameters in this model.
Another, perhaps circumstantial, evidence of SUSY is the
successful unification of the gauge couplings at some high scale.
Add to this the fact that the neutralino can provide a good dark
matter candidate explains the popularity of the model. Even so the
search for the lightest Higgs is not so easy. LEP2 where the Higgs
signature is easiest may unfortunately be some $20-30$~GeV short
to be able to cover the full range of the minimal SUSY lightest
Higgs mass. Searches at the Tevatron need very good background
rejection and in any case need to upgrade the present luminosities
quite significantly. At the LHC, most analyses have relied
extensively on the two-photon decay of the IMH either in the
dominant inclusive channel through $gg \ra h \ra \gamma \gamma$ or
in associated production. Only recently has it been shown that
associated production of the Higgs with tops with the former
decaying into $b \bar b$ can improve the discovery of the Higgs,
albeit in the region $m_h <120$GeV\cite{ATLAS_TDR}. Unfortunately,
until recently\cite{ATLAS_TDR}, most simulations for Higgs
searches have in effect decoupled the rest of the supersymmetric
spectrum from the Higgs sector, like in the much advertised
ATLAS/CMS $M_A-\tgb$ plane\cite{ATLAS_TDR,CMS_allHiggs}.

This assumption of a very heavy SUSY spectrum can not be well
justified. First, naturalness arguments require that at least some
of the SUSY masses be below $1TeV$ or even much less. Second, it
has been known\cite{Kileng_mixing,RggKane} that relaxing this
assumption can have some very important consequences on the Higgs
search at the LHC. This is not surprising considering the fact
that the most important production channel $gg \ra h$ is loop
induced as is the main discovery channel $h \ra \gagt$. One of the
most dramatic effect is that of a light stop with large mixing
which drastically reduces the production
rate\cite{Kileng_mixing,AbdelStop_Hgg_Loops}. Fortunately, when
this happens, a careful analysis\cite{nous_Rggstophiggs_lhc} shows
that the Higgs signal can be rescued in a variety of channels that
become enhanced or that open up precisely for the same reason that
the normal inclusive channel drops, so that in a sense there is a
complementarity.  For instance with all other sparticles but the
stops heavy, one can show that whenever the production rate in the
inclusive channel drops, the branching ratio into two photons
increases with the consequence that associated $Wh/Zh$ and $t\bar
t h$ where  the Higgs decays into two photons becomes a very
efficient means of tracking the Higgs. Moreover associated $\sto
\sto h$
production\cite{nous_Rggstophiggs_lhc,stophiggs_LHC,Moretti_stophiggs}
becomes important  through the cascade of the heavier stop $\stt$,
$\stt \ra \sto h$. At the same time since the $h b \bar b$
coupling is hardly affected $t \bar t h$ production could play an
important role. Similar sort of complementarity has also been
pointed out in supersymmetric scenarios where the coupling $h b
\bar b$ can be made very small\cite{CWMrenna-Tevatron}.

In our investigation of the effects of light stops with large
mixing, all other particles but the stops were assumed rather
heavy. It is then important to ask how the overall picture changes
had we allowed other  sparticles to be relatively light.
Considering that the present LEP and Tevatron data precludes the
decay of the lightest Higgs into sfermions, the effect of the
latter on the properties of the lightest Higgs can only be felt
through loops. These effects can therefore be considered as a
special case of the stop that we studied at some length and apart
from the sbottom at large $\tgb$ the effects will be marginal. One
can then concentrate on the spin-half  gaugino-higgsino sector. In
order to extract the salient features that may have an important
impact on the Higgs search at the LHC, we leave out in this study
the added effects of a light stop. Compared to the analysis with
the stop, this sector does not affect inclusive production nor the
usual associated production mechanisms. The effect will be limited
to the Higgs decay. First, if the charginos are not too heavy they
can contribute at the loop level. We find however, by imposing the
present limits on their masses, that this effect is quite small.
On the other hand we show that the main effect is due to the
possible decay of the Higgs into the lightest neutralino. This is
especially true if one relaxes the usual so-called unification
condition between the two gaugino components of the LSP
neutralino. Although at LEP an invisible Higgs is not so much of a
problem\cite{aleph-hinvisible}, since it can be easily tagged
through the recoiling Z, it is a different matter at the LHC. Few
studies have attempted to dig out such an invisible (not
necessarily supersymmetric) Higgs at the LHC, in the associated
$Zh, (Wh)$\cite{Kane-invisible-lhc, DP-invisible-lhc} channel.
Even with rather  optimistic rejection efficiencies  the
backgrounds seem too overwhelming. It has also been suggested
\cite{Gunion-invisible-lhc} to use associated $t \bar t h$
production but this requires very good $b$-tagging efficiencies
and a good normalisation of the backgrounds. Recently
Ref.~\cite{Tevatron-invisible-lhc} looked at how to hunt for an
invisible Higgs at the Tevatron. For $m_h>100$GeV a $5\sigma$
discovery requires a luminosity in excess of $30$fb$^{-1}$.

 Compared to the effects of the stop or a vanishing
$ h b \bar b$, where a sort of compensation occurs in other
channels, the opening up of the invisible decay reduces all other
channels including the branching ratio in $b \bar b$. Previous
studies\cite{Haber-old-invisible,Abdel-h-invisible} have mainly
concentrated if not on a mSUGRA scenario then on a scenario based
on the mSUGRA inspired relation between the electroweak gaugino
masses, $M_1,M_2$. Moreover LEP searches and limits refer
essentially to the latter paradigm. In the course of this analysis
we had to re-interpret the LEP data in the light of more general
scenarios. We therefore had to take recourse to various limits on
cross sections rather than absolute limits on some physical
parameters quoted in the literature. We have also tried to see
whether new mechanisms come to the rescue of the Higgs search at
the LHC when the invisible channel becomes substantial and reduces
the usual signal significantly. Much like in our analysis of the
stop\cite{nous_Rggstophiggs_lhc} where we found that the Higgs
could be produced through the decay of the heavier stop into the
lighter one, we inquired whether a similar cascade decay from the
heavier neutralino or charginos to their lighter companions can
take place. This is known to occur for instance for some mSUGRA
points\cite{gluinoscascadetoh,ATLAS_TDR}, but its rate is found
not to be substantial when an important branching ratio into
invisibles occurs. Even if it were substantial it would be
difficult to reconstruct the Higgs since again at the end of the
chain the Higgs will decay predominantly invisibly.

Considering the dire effects of a large invisible branching ratio
occurring for rather light neutralinos, we have investigated the
astrophysical consequences of such scenarios, specifically the
contribution of such light neutralinos on the relic density. We
find that these models require  rather light slepton masses. In
turn, with such light sleptons, neutralino production at LEP2
provides much restrictive constraints than the chargino cross
sections. Taking into account the latter constraints helps rescue
some of the Higgs signals.

The paper is organised as follows. In the next section we
introduce our notation for the chargino-neutralino sector and make
some qualitative remarks concerning the coupling of the lightest
Higgs as well as that of the $Z$ to this sector.  This will help
understand some of the features of our results. In section 2 we
review the experimental constraint and discuss how these are to be
interpreted within a general supersymmetric model. Section 3
presents the results for Higgs detection at the LHC within the
assumption of the GUT relation for the gaugino masses. Section 4
analyses the ``pathological" cases with a sneutrino almost
degenerate with the chargino, leading to lower bounds on the
chargino mass. Section 5 analyses how the picture changes when one
relaxes the GUT inspired gaugino masses constraint and the impact
of  the  astrophysical constraints on the models that may
jeopardise the Higgs search at the LHC. Section 6 summarises our
analysis.

\section{The physical parameters and the constraints}
\subsection{Physical parameters}
When discussing the physics of charginos and neutralinos it is
best to start by defining one's notations and conventions. All our
parameters are defined at the electroweak scale. The chargino mass
matrix in the gaugino-higgsino basis is defined as
\beqn
\label{charginomatrix} \left(
\begin{array}{cc}
M_2 & \sqrt{2} \mw \cosb  \\ \sqrt{2} \mw \sinb   & \mu
\end{array}  \right)
\eeqn

\noindent where $M_2$ is the soft SUSY breaking mass term for the $SU(2)$
gaugino while $\mu$ is the so-called higgsino mass parameter
whereas $\tgb$ is the ratio of the vacuum expectation values for
the up and down Higgs fields.

Likewise the neutralino mass matrix is defined as

\beqn
\label{neutralinomatrix} \left(
\begin{array}{cccc}
M_1 & 0 & -\mz \sw \cosb & \mz \sw \sinb \\ 0 & M_2 & \mz \cw
\cosb & -\mz \cw \sinb \\ -\mz \sw \cosb   & \mz \cw \cosb  & 0 &
-\mu \\ \mz \sw \sinb & -\mz \cw \sinb & -\mu & 0
\end{array}  \right)
\eeqn

\noi where the first entry $M_1$ (corresponding to the bino
component) is the $U(1)$ gaugino mass. The oft-used gaugino mass
unification condition corresponds to the  assumption

\beqn
\label{m1m1unification} M_1=\frac{5}{3} \tan^2\theta_W M_2 \simeq
M_2/2
\eeqn

Then constraints from the charginos alone can be easily translated
into constraints on the neutralino sector. Relaxing
Eq.~\ref{m1m1unification}, or removing any relation between $M_1$
and $M_2$ means that one needs further observables specific to the
neutralino sector.

The other parameters that appear in our analysis emerge from the
Higgs sector. We base our study on the results and prescription of
\cite{CarenaWagner_Higgs_Approx1} for the improved two-loop
calculations based on the effective Lagrangian\footnote{There is
now a two-loop diagrammatic calculation\cite{FeynHiggs} which is
in  good agreement with an updated version of the two-loop
effective Lagrangian approach
\cite{mh_RGEvsExact,Espinosa-Zhang99}.}. The parameters here are,
apart from the ubiquitous $\tgb$, the mass of the pseudo-scalar
Higgs, $M_A$, $A_t$ the trilinear mixing parameter in the stop
sector, as well as $M_S$ a supersymmetric scale that may be
associated to the scale in the stop sector. Since we want to
delimit the problem compared to our previous study on the stop
effects, we will set the stop masses (and all other squarks) to
$1$TeV. We will also be working in the decoupling limit of large
$M_A$ that we also set at 1TeV. The lightest Higgs mass is then
larger than if we had taken a lower $M_A$. As we will see the most
important effect that results in small branching ratio for the
two-photon width is when the invisible decay opens. This occurs if
one has enough phase space and therefore if the mass of the Higgs
is made as large as possible. Thus for a given $\tgb$ the effect
is maximal for what is called maximal mixing: $A_t \sim \sqrt 6
M_S$ in the implementation of \cite{CarenaWagner_Higgs_Approx1}.
One would also think that one should make $\tgb$ large, however
this parameter also controls the masses of the neutralinos and for
the configuration of interest, those leading to the largest drops
in the two-photon signal, one needs to keep $\tgb$ as low as
possible to have the lightest neutralino as light as possible.

In principle we would have liked to decouple all other sparticles,
specifically sfermions as stated in the introduction. However
sleptons (in particular selectrons and sneutrinos) masses
determine also the cross sections and the decay signature of the
charginos and the neutralinos. Therefore, allowing for smaller
sfermions masses does not so much directly affect the two-photon
width but can relax quite a bit some of the limits on the
chargino-neutralino sector which in turn affect the Higgs search.
We thus allow for this kind of indirect dependence on the sfermion
mass.

Often, especially in the case of neutralinos, LEP analyses set
{\em absolute} bounds  on masses. Ideally, since one is using
bounds that are essentially set from the couplings of neutralinos
to gauge bosons, to translate to couplings of these neutralinos
and charginos to the Higgs, one needs to have access to the full
parameter space $\mu, \tgb, M_1,M_2$. Thus absolute bounds are
only indicative and it is much more informative to reinterpret the
data. In the case of limits set solely from the chargino data, the
re-interpretation is quite straightforward since no  assumption on
the parameters in Eq.~\ref{charginomatrix} is made and the limits
ensue from $\epem \ra \chargop \chargom$. Limits on the
neutralinos are a bit more involved. To make some of these points
clearer and to help understand some of our results it is worth
reviewing the couplings to neutralinos.

\subsection{Couplings of Neutralinos to the Higgs and $Z$}
The width of the lightest Higgs to the lightest neutralinos
writes\cite{Haber-old-invisible}

\beqn
\label{gamhtolsp} \Gamma(h \ra \neuto \neuto)=\frac{G_F \mw \mh}{2
\sqrt{2} \pi} \;(1-4 \mneuto^2/m_h^2)^{3/2}\;  |C_{h \neuto
\neuto}|^2
\eeqn

\noindent where\cite{HHG}
\beqn
\label{widthhtoneutalino} C_{h \neuto \neuto}&=&(O^N_{12}-\tw
O^N_{11})(\sinal\; O^N_{13}\;+\; \cosal \; O^N_{14}) \nonumber
\\  &\simeq&
(O^N_{12}-\tw O^N_{11})(\sinb\; O^N_{14}\;- \; \cosb \;O^N_{13})
\eeqn
$O^N_{ij}$ are the elements of the orthogonal ( we assume \cpviol
conservation) matrix which diagonalizes the neutralino mass
matrix. $\alpha$ is the angle that enters the diagonalization of
the CP-even neutral Higgses which in the decoupling (large $M_A$
and ignoring radiative corrections) is trivially related to the
angle $\beta$. $|O^{N}_{1j}|^2$ defines the composition of the
lightest neutralino $\neuto$. For instance $|O^{N}_{11}|^2$ is the
bino purity and $|O^{N}_{11}|^2+|O^{N}_{12}|^2$ is the gaugino
purity. It is clear then, apart from phase space, that the LSP has
to be a mixture of gaugino and higgsino in order to have a large
enough coupling to the Higgs. The same applies for the diagonal
coupling of the charginos ($h \chi_i^- \chi_i^+$).

\begin{figure*}[htbp]
\begin{center}
\mbox{
\includegraphics[width=8cm,height=8cm]{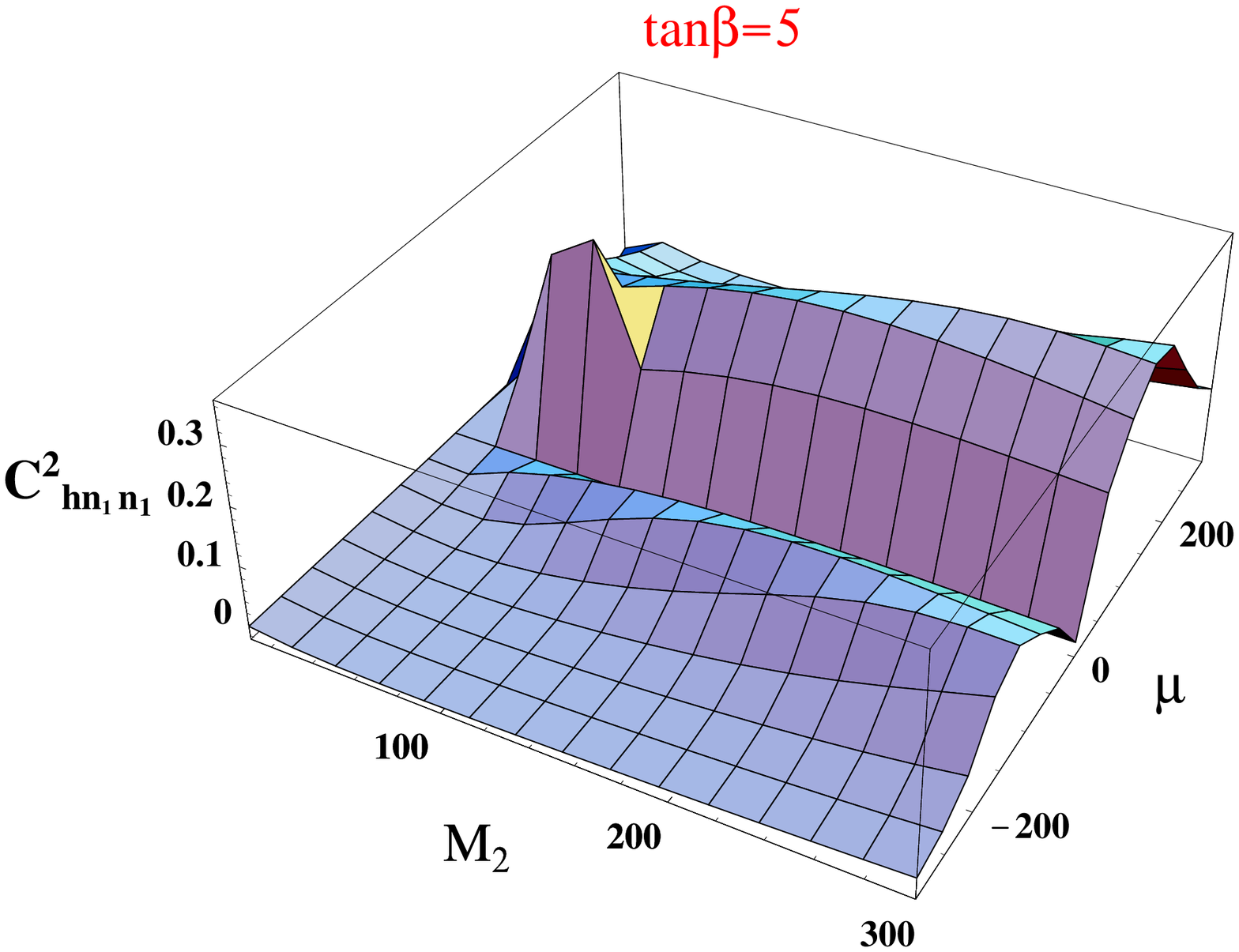}
\includegraphics[width=8cm,height=8cm]{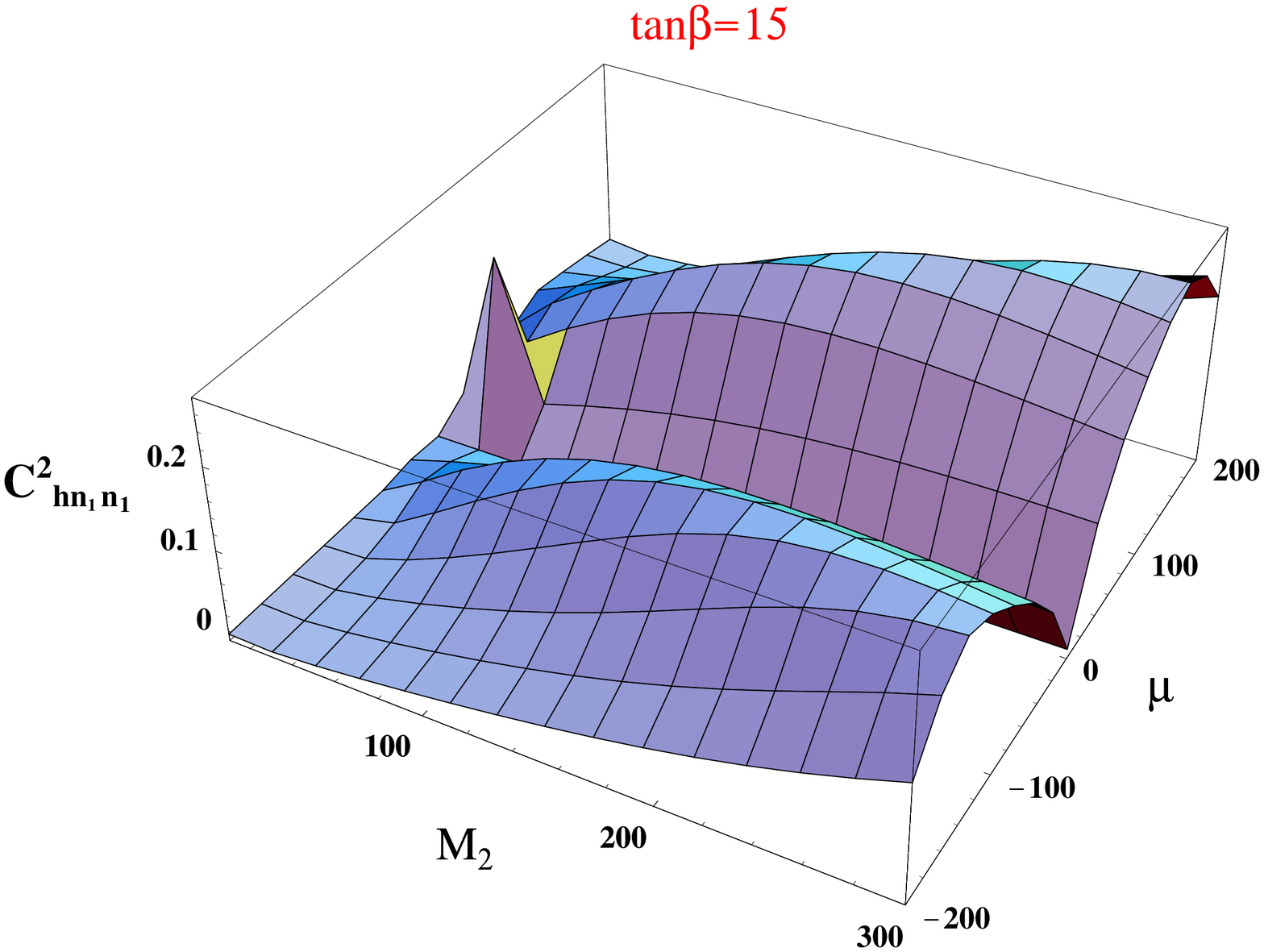}}
\mbox{
\includegraphics[width=8cm,height=8cm]{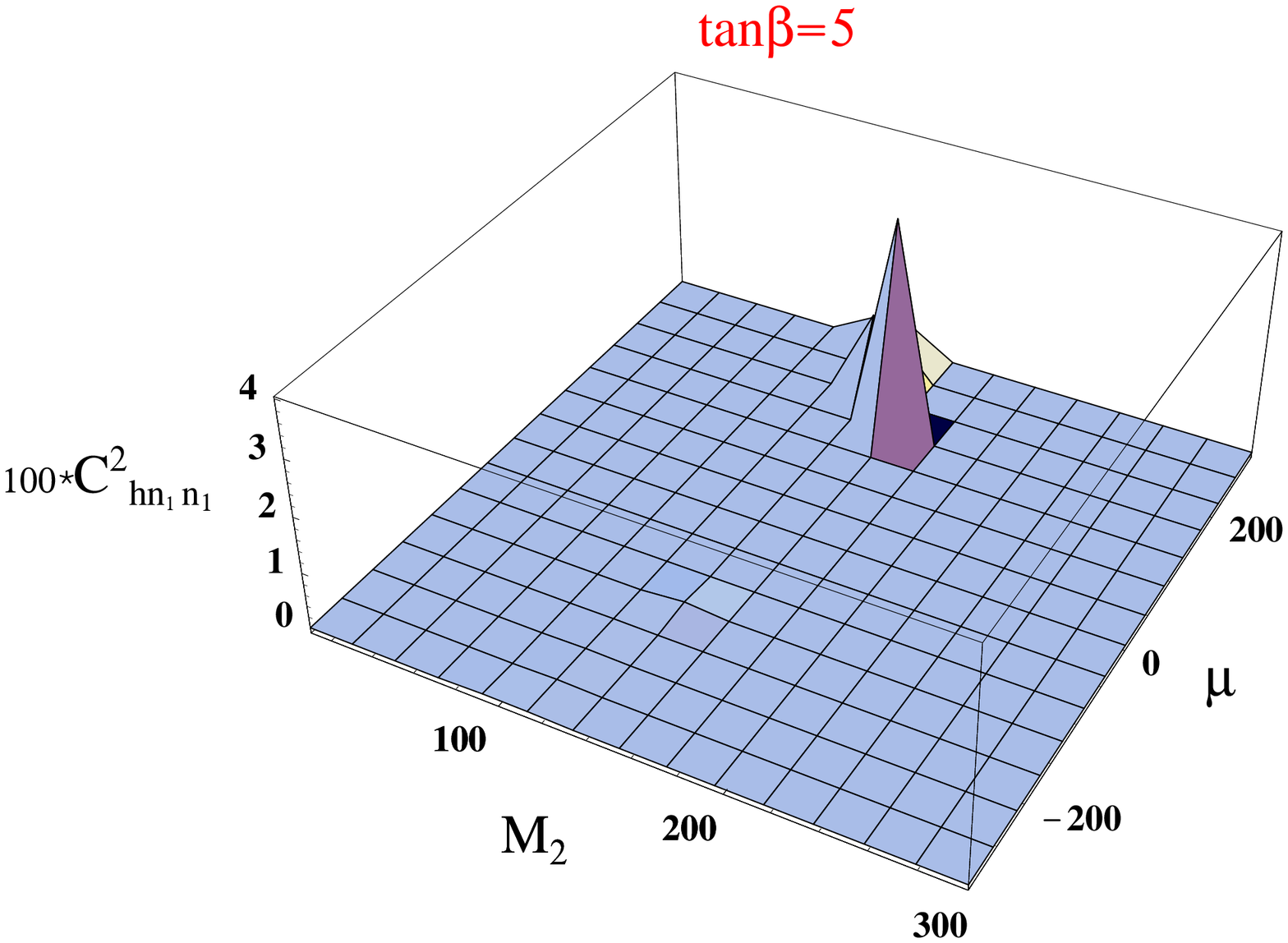}
\includegraphics[width=8cm,height=8cm]{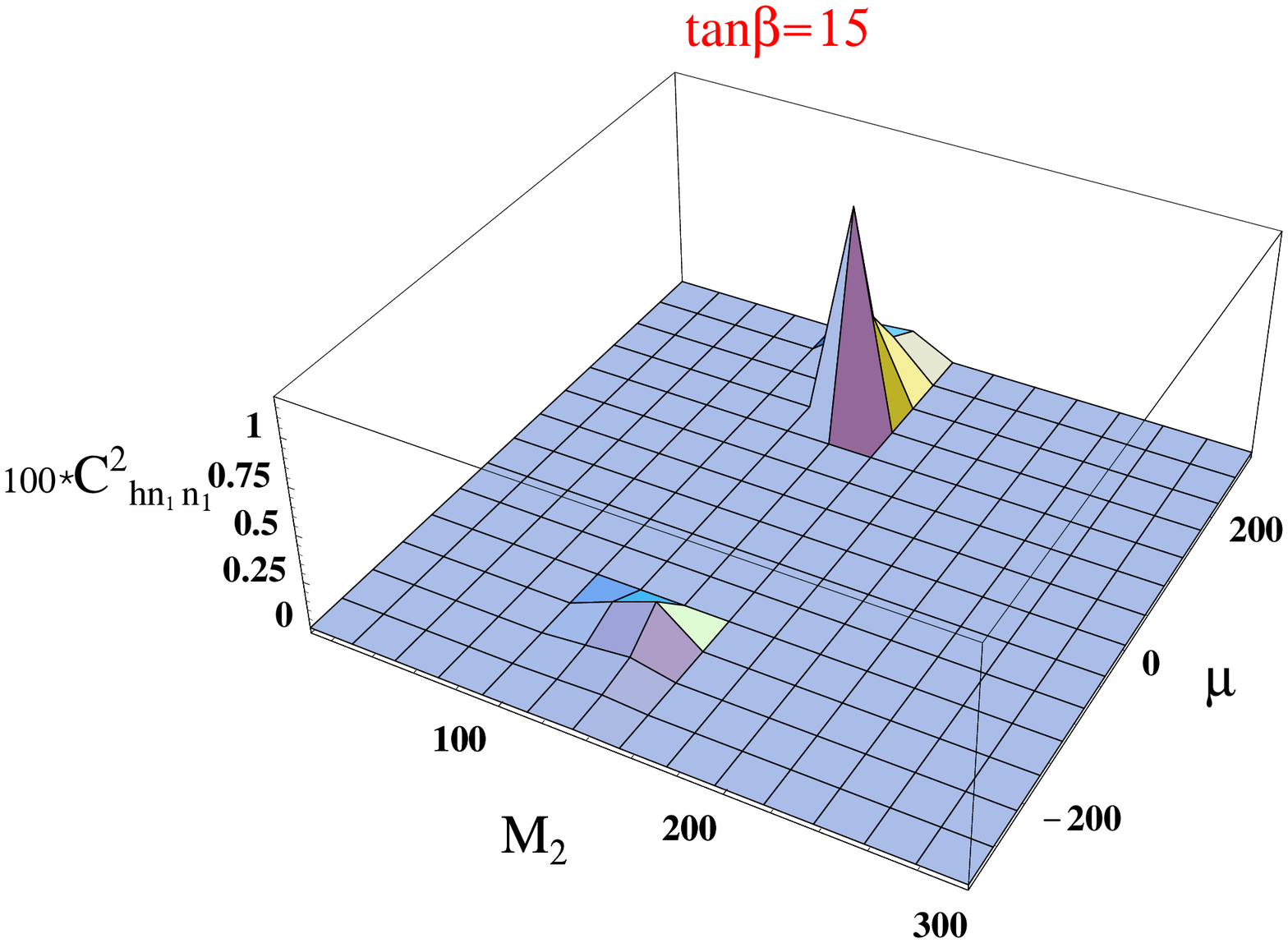}}
\caption{\label{chn1n1sqtgb5}{\em Strength of the lightest Higgs
coupling to $\neuto \neuto$ for $\tgb=5$ and $\tgb=15$. In the
second set (bottom) of figures we have imposed $\mchargo >
94.5$GeV (the current limit on the lightest chargino) and $\mneuto
< 65$GeV (this corresponds to the threshold for the lightest Higgs
with a maximum mass of about $130$GeV to decay into the LSP
neutralino). Here $C_{hn_1n_1} \equiv C_{h\neuto \neuto}$ \/.}}
\end{center}
\end{figure*}

\begin{figure*}[htbp]
\begin{center}
\mbox{
\includegraphics[width=8cm,height=8cm]{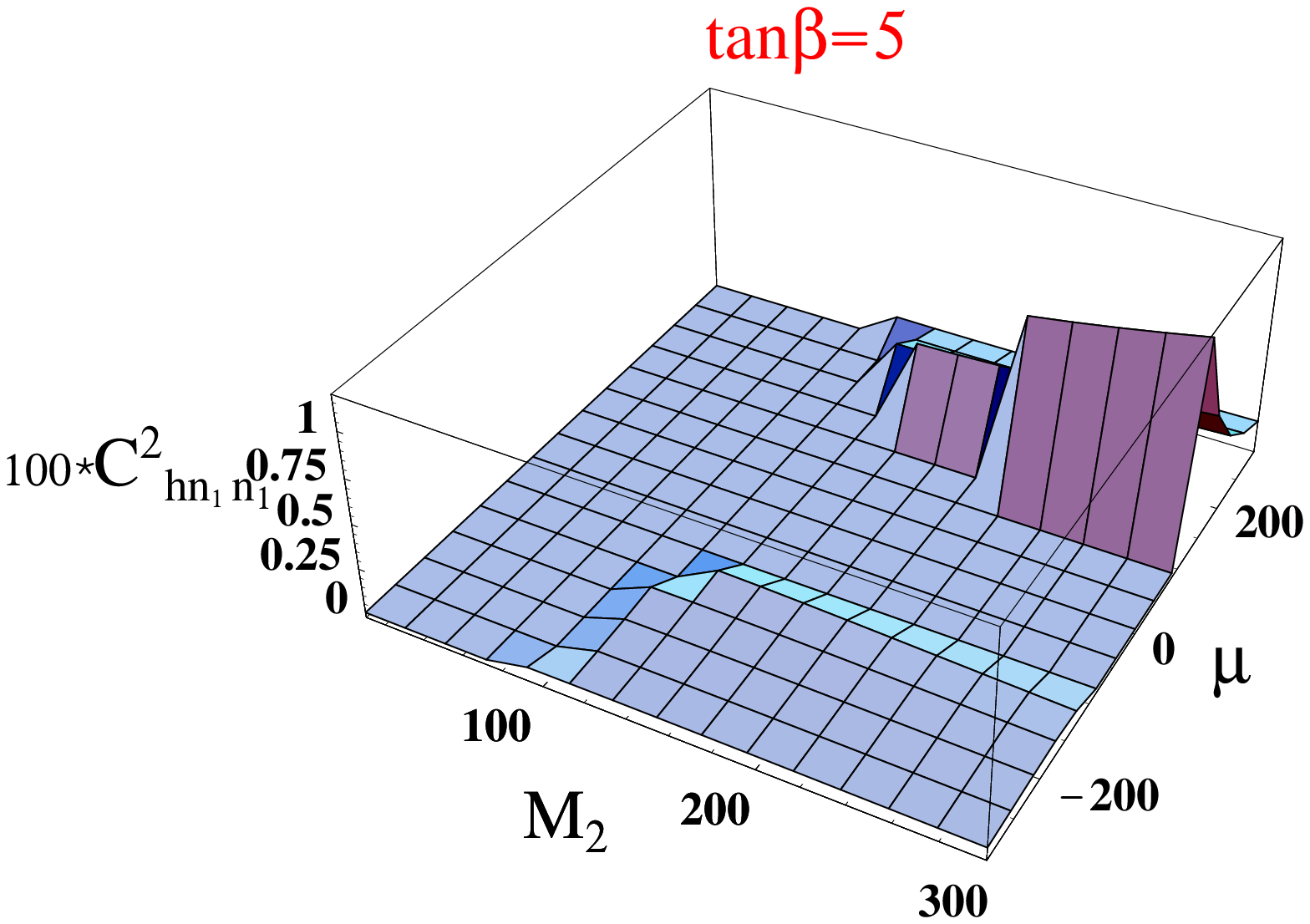}
\includegraphics[width=8cm,height=8cm]{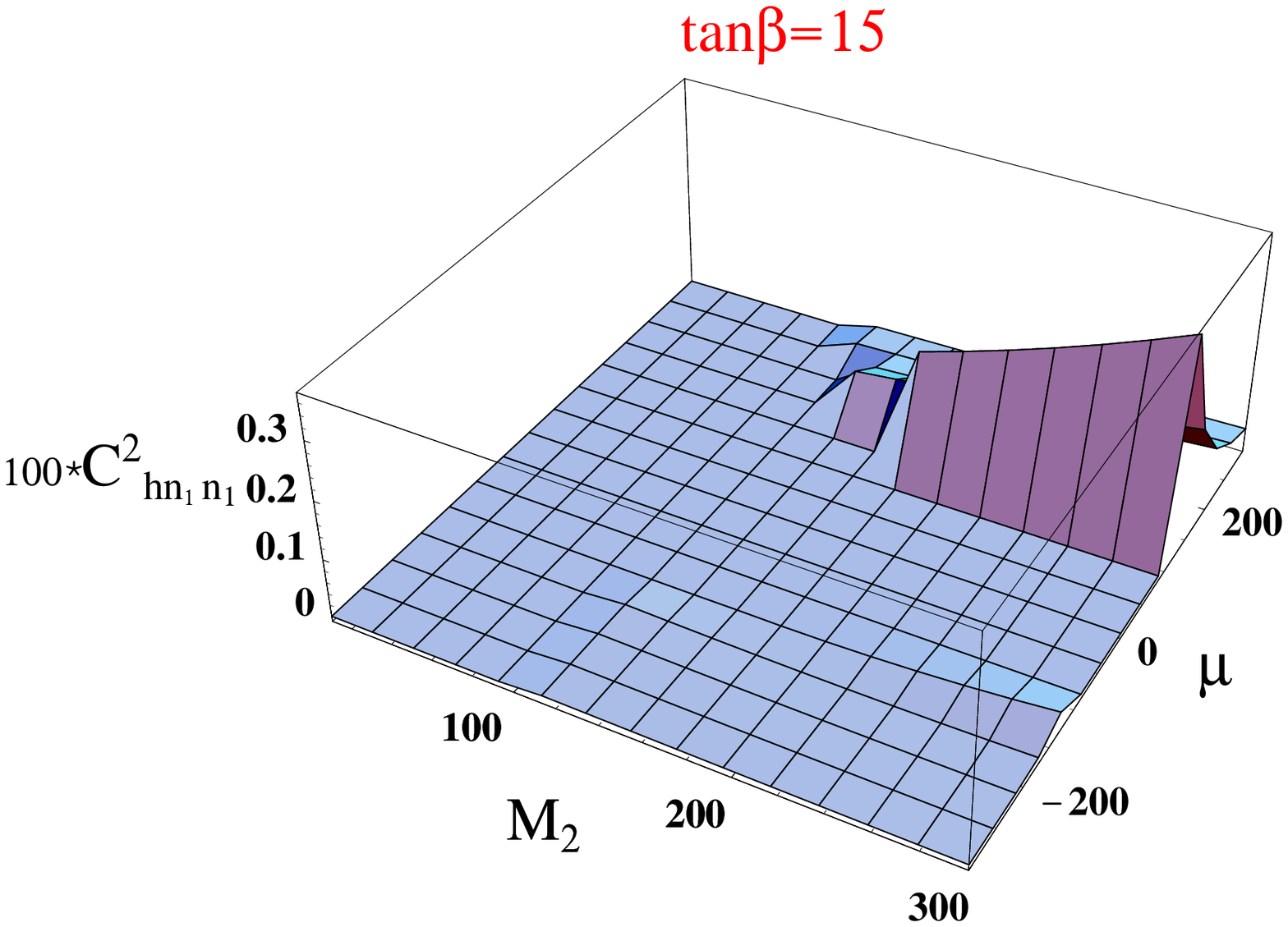}}
\caption{\label{chn1n1sq_nouni}{\em Strength of the lightest Higgs
coupling to $\neuto \neuto$ for $\tgb=5$ and $\tgb=15$ in the case
$M_1=M_2/10$ . As in Fig.~\ref{chn1n1sqtgb5}, we have imposed
$\mchargo > 94.5$GeV and $\mneuto
< 65$GeV\/.}}
\end{center}
\end{figure*}

In Fig.~\ref{chn1n1sqtgb5} we show the strength $C_{h \neuto
\neuto}^2$ assuming the GUT unification condition between $M_1$
and $M_2$ for $\tgb=5$ and $\tgb=15$. One should note that the
coupling is much larger  for positive values of $\mu$. The largest
effect (peak) occurs for small values of $\mu$ and $M_2$ which
however are ruled out by LEP data on the chargino mass. Note also,
by comparing the $\tgb=5$ and $\tgb=15$ case in
Fig.~\ref{chn1n1sqtgb5}, that especially for $\mu >0$, as \tgbt
increases the Higgs coupling to the LSP gets smaller. At the same
time  the neutralino LSP gets heavier. Thus large \tgbt values
corresponding to higher Higgs masses will not lead to the largest
$h\ra \neuto \neuto$.
Similar behaviour is also observed for the coupling of the
chargino to Higgs, the largest coupling sits in the $\mu > 0$ and
small $M_2$ region. However, it turns out that the effect of
charginos in the loop never becomes very large.
\\ \noindent As
seen in Fig.~\ref{chn1n1sq_nouni} for the case with $M_1=M_2/10$,
the same kind of behaviour persists:  $\mu>0$ and moderate \tgbt
lead to stronger couplings.
\noindent On the other hand, the constraints on $M_2, M_1,\mu$
which are derived for instance from neutralino production are more
sensitive to the higgsino component of the neutralino. Indeed the
$Z$ coupling  to these writes
\beqn
\label{Ztoneutralino} Z^* \ra \chi^0_i \chi^0_j \propto (O^N_{i3}
O^N_{j3} -\; O^N_{i4} O^N_{j4} )^2
\eeqn

\noi Chargino production in \epemt is not as much critically
dependent on the amount of mixing since both the wino and
(charged) higgsino components couple to the $Z$ and the photon.
Some interference with the t-channel sneutrino exchange may occur
in the case of a wino component ({\it i.e.} $|\mu| \ll M_2$),
therefore the kinematic limit can be reached quite easily, except
the situation where the signature of the chargino leads to almost
invisible decay products.

\subsection{Accelerator Constraints}
This brings us to how we have set the constraints.

$\bullet$ \underline{Higgs mass}:\\ \noi  In the scenarios we are
considering with large $M_A$ and large stop masses the $ZZh$
coupling is essentially SM-like and LEP2 limits on the mass of the
SM Higgs apply with little change even for an invisible Higgs. In
any case, as discussed earlier, to make the chargino-neutralino
effect most dramatic we will always try to maximise the Higgs mass
independently of $\tgb$ by choosing an appropriate $A_t$. The LEP2
mass limit are thus always evaded. For $M_1=M_2/10$ we stick to
$\tgb=5$, considering there is always enough phase space for $h\ra
\neuto \neuto$, it is sufficient to discuss the case with $A_t=0$.
For the canonical unification case, the effect of maximising the
Higgs mass through $A_t$ is crucial.

$\bullet$ \underline{Chargino cross section}:
\\ Typically when no
sparticle is degenerate with the chargino, the lower limit on the
chargino mass reaches the LEP2 kinematic limit independently of
the exact composition of the chargino and does not depend much on
the  sneutrino mass as explained earlier. Latest LEP data give
\cite{L399chiplus},
\beqn
\label{mchi90} m_{\chi_1^+} \geq 94.5GeV.
\eeqn
Very recent combined  preliminary data\cite{mh_limit-nov99}
suggest $m_{\chi_1^+} \geq 100.5GeV$. We will also comment on how
our results can change by imposing this latter limit.

\hspace*{1cm} \underline{{\em Degeneracy with the LSP}}
\\ Even when slepton masses can be large, in which case the chargino cross
section is larger, the chargino mass constraint weakens by a few
GeV when the lightest chargino and neutralino are almost
degenerate\cite{L399chiplus}. The $\chargop \ra \neuto f \bar f'$
decay leads to soft ``visible" products that are difficult to
detect . Recent LEP data has greatly improved the limits in this
small $\Delta M_{\chargop-\neuto}$ mass difference region. However
within the assumption of gaugino mass unification the highly
degenerate case with  a light chargino/neutralino occurs in the
region $\mu \ll M_2, M_2 \geq 2$~TeV. In this region the light
(and degenerate) neutralino and chargino are almost purely
Higgsino and therefore as seen from Eq.~\ref{widthhtoneutalino} do
not couple strongly to the Higgs. Their effect on the Higgs
invisible width as well as indirectly on the  two-photon width is
negligible. We will not consider this case.

\hspace*{1cm} \underline{{\em Degeneracy with a light sneutrino}}:
\\There is another degeneracy which is of more concern to
us. It occurs for small slepton masses that are almost degenerate
with the chargino, rendering the dominant two-body decay mode
$\chi^+\ra \tilde\nu l^+$ undetectable (the three flavours of
sneutrinos are also degenerate).  When this occurs, for
$\Delta_{\rm deg}=\mchargo -\msnue<3GeV$, neutralino production is
also of no use since the neutralinos will also decay into
invisible sneutrinos. Since SU(2) relates the mass of the
sneutrinos to that of the left selectrons, the search for the
latter will then set a limit on the charginos in this scenario.
 The explorable mass of the selectron is a few GeV from the
LEP2 kinematical limit. In fact the LEP Collaborations make a
stronger assumption to relate the mass of the sneutrinos to those
of the selectrons. Left and right sleptons masses are calculated
according to a mSUGRA scenario  by taking a common scalar mass,
$m_0$,  defined at the GUT scale. This gives
\beqn
\label{m0sugra} \mser^2&=&m_0^2\;+\;0.15 M_{1/2}^2\;-\;\sww D_z
\nonumber
\\ \msel^2&=&m_0^2\;+\;0.52 M_{1/2}^2\; -\;(.5-\sww)D_z \nonumber
\\ \msnue^2&=&m_0^2\;+\;0.52 M_{1/2}^2\;+\;D_z/2  \;\;\;\;\;\; {\rm with}
\nonumber
\\  D_z&=&\mzz \cosbb
\eeqn
where $M_{1/2}$ is the common gaugino mass at the GUT scale also,
which we can relate to the $SU(2)$ gaugino mass as $M_2 \sim 0.825
M_{1/2}$. With these assumptions, $\mser$ gives the best limit.
One thus arrives at a limit\cite{L399chiplus}
\beqn
\label{mchi70} m_{\chi_1^+}\simeq \msnue \geq 70GeV (\tgb=5).
\eeqn

 The above reduction in the chargino mass limit compared to Eq.~\ref{mchi90}
will have dramatic effects on the Higgs two-photon width. In this
very contrived scenario the conclusions we will reach differ
significantly from the general case. This very contrived scenario
will be discussed separately in section 4.

$\bullet$ \underline{{\em Neutralino Production and decays}}:
\\
LEP2  also provides a constraint on the mass of the  neutralino
LSP from the search for a pair of  neutralinos, specifically
$e^+e^-\ra \chi^0_i\chi^0_j$. This constraint is relevant for the
small $(\mu, M_2)$ and also when we relax the unification
condition. We have implemented the neutralino constraint by
comparing the crosssection for neutralino production with the
tables containing the upper limit on the production cross-section
for $\chi_1^0\chi_j^0$ obtained by the L3 collaboration at
$\sqrt{s}=189GeV$\cite{L3neutralino99}. These tables give an upper
limit on the cross-section for the full range of kinematically
accessible $\neuto+\neutt$ masses. The limits depend in a
non-trivial manner on the masses of the produced particles.
Moreover the limits are slightly different depending on whether
one assumes purely hadronic final states from the decay of the
heavier neutralino or whether one assumes leptonic final states.
\footnote{The combination of the various selections, the leptonic
and hadronic ones, is an {\it a priori} optimisation using Monte
Carlo signal and background events. This optimisation procedure is
defined to maximise the signal efficiency (including the leptonic
and hadronic branching ratios) and minimise the background
contamination. This consists in the minimisation of $\kappa$,
expressed mathematically by: $\kappa=\Sigma_{n=0}^{\infty}
k(b)_{n} P(b,n)/\epsilon$, where $k(b)_{n}$ is the $95\%$
confidence level Bayesian upper limit, $P(b,n)$ is the Poisson
distribution for $n$ events with an expected background of $b$
events, and $\epsilon$ is the signal efficiency including the
branching ratios.}
 Under the same assumptions we have also used these
tables for setting the upper limit on $\neuto \neutth$ production.
In all models where gaugino mass unification is imposed, the
virtual Z decay mode, $\chi^0_{2,3} \ra \neuto Z^*$, constitutes
the main decay mode when the neutralinos are light enough to be
accessible at LEP2. In models where the gaugino mass unification
is relaxed and very light neutralinos exist, as will be discussed
in section ``no-unification", other decay channels may open up,
for example $\neutth \ra\neuto h$. The analysis, and hence the
derived constraints, is made more complicated if one allows for
light sleptons as will be suggested by cosmology in these models.
Though light sleptons enhance the neutralino cross section quite
significantly, in the case of left sleptons the efficiency is
degraded because the branching ratio of the heavier neutralino
into invisible (through a three body or even two-body $\neutt \ra
\nu \snu^*$) may be important.

As just discussed one also needs to take into account the various
branching ratios of the neutralinos and charginos. These were also
needed when considering production of neutralinos and charginos at
the LHC.  We have taken into account all two-body and three-body
decay modes of gauginos , including fermionic and bosonic final
states, $\chi^0_j\ra\chi^0_i Z,\chi^0_i h,\chi^{\pm}W^{\mp},
\tilde{l}l$ and $\chi^+_j\ra\chi^+_i Z,\chi^+_1 h,\chi^0_j W^{+},
\tilde{l}l, \tilde{\nu}\nu$ . The analytical formulas were checked
against the outputs of programs for automatic calculations such as
{\tt GRACE} \cite{grace} and {\tt COMPHEP}\cite{comphep} . For
channels involving a Higgs boson, the radiatively corrected Higgs
mass as well as the Higgs couplings, $\sin\alpha$, following the
same implementation as  in \cite{HDECAY} were used.

$\bullet$ \underline{{\em Invisible width of the $Z$ and single
photon cross section at LEP2}}:
\\
In the case were we lift the unification condition that leads to
rather small neutralino masses which are kinematically accessed
through $Z$ decays we have imposed the limits on the invisible
width of the $Z$\cite{LEPelectroweak}:
\beqn
\label{invisibleZ} \Gamma^Z_{{\rm inv}}\equiv \Gamma(Z\ra \neuto
\neuto) < 3 MeV
\eeqn
In view of the limits on the single photon cross section which can
translate into limits on $\sigma(\epem \ra \neuto \neuto \gamma)$,
with cuts on the photon such that $E_\gamma>5GeV$ and
$\theta_{{\rm beam}-\gamma}> 10^0$, we used $\sigma(\epem \ra
\neuto \neuto \gamma)< .1pb$ at $\sqrt{s}=189GeV$. In fact L3
gives a limit of $.3pb$\cite{L3radiativeLSP99}, foreseeing that
similar analysis will be performed for the other collaborations
and the results will be combined we conservatively took $.1$pb.
However this constraint turned out not to be of much help.

\subsection{Cosmological  Constraints}
Scenarios with $M_1=M_2/10$ have very light neutralino LSP into
which the Higgs can decay, suppressing quite strongly its visible
modes. Accelerator limits still allow for such a possibility.
However, it has been known that  a very light neutralino LSP can
contribute quite substantially to the relic density if all
sfermions are light. In the last few  years constraints on the
cosmological parameters that enter the calculation of the relic
density have improved substantially. Various
observations\cite{omega} suggest to take as a benchmark $
\Omega_\chi h_0^2 <.3$ where we identify $\Omega_\chi$ with  the
fraction of the critical energy density provided by neutralinos.
$h_0$ is the Hubble constant in units of 100 km sec$^{-1}$
Mpc$^{-1}$. This constraint is quite consistent with limits on the
age of the Universe\cite{ageuniverse}, the measurements of $h_0$ 
\cite{hubble}, the measurements of the lower multipole moment 
power spectrum from CMB data and the determination of $\Omega_{\rm 
matter}$ from rich clusters, see \cite{omega} for reviews. It 
also, {\em independently}, supports data from type Ia 
supernovae\cite{supernova} indicative of a cosmological constant. 
For illustrative purposes and to show how sensitive one is to this 
constrain we will also consider a higher value, $ \Omega_\chi 
h_0^2 <.6$ that may be entertained if one relies on some mild 
assumptions based on the age of the Universe and the CMB result 
only\cite{wells-cosmo}. In this scenario the calculation of the 
relic density is rather simple since one only has to take into 
account annihilations into the lightest fermions. Keeping with our 
analysis, we required all squarks to be heavy but allowed the 
sleptons to be light. To calculate the relic abundance we have 
relied on a code whose characteristics are outlined in 
\cite{torino-dm}. To help with the discussion we will also give an 
approximate formula that agrees better than $30\%$ with the 
results of the full calculation. 

\subsection{LHC Observables}
The principal observables we are interested in are those related
to the Higgs production and decay. Since we are only considering
the effects of non-coloured particles and are in a regime of large
$M_A$, all the usual production mechanisms (inclusive direct
production through gluon-gluon as well as the associated
production $W(Z) h$ and $t \bar{t} h$) are hardly affected
compared to a SM Higgs with the same mass. Contrary to the
indirect effects of light stops and/or sbottoms, the main effects
we study in this paper affect only  decays of the Higgs. The main
signature into photons is affected both by the indirect loop
effects of light enough charginos (and in some cases sleptons) and
by the possible opening up of the Higgs decay into neutralinos.
When the latter is open it leads to the most drastic effects
reducing both the branching into photons as well as into $b \bar
b$, hence posing a problem even for the search in $t \bar t h$
with $h \ra b \bar b$. To quantify the changes of the branching
ratios we define, as in \cite{nous_Rggstophiggs_lhc}, the ratio of
the Higgs branching ratio into photons normalised to that of the
SM, defined for the {\em same value of the Higgs mass}:
\beqn \label{Rgamgam}
 R_{\gamma\gamma}=\frac{ BR^{SUSY}(h
\ra \gamma \gamma)}{BR^{SM}(h \ra \gamma \gamma)}
\eeqn
Likewise for the branching ratio into $b \bar b$
\beqn \label{Rbb}
 R_{b \bar b}=\frac{ BR^{SUSY}(h
\ra b \bar b)}{BR^{SM}(h \ra b \bar b)}
\eeqn

The latter signature for the Higgs has only recently been analysed
within a full ATLAS simulation and found to be very useful for
associated $t \bar t h$ production, but only for
$m_h<120$GeV\cite{ATLAS_TDR}. With $100$fb$^{-1}$ the significance
for a \sm Higgs with $m_h=100$GeV is 6.4 but drops to only $3.9$
for $m_h=120$GeV. Since this is the range of Higgs masses that
will interest us, we will consider a drop corresponding to
$R_{b\bar b}=.7$ to mean a loss of this signal.

As concerns the two-photon signal, we take $R_{\gamma \gamma}<.6$
as a benchmark for this range of Higgs masses. This is  somehow a
middle-of-the-road value between the significances given by
ATLAS\cite{ATLAS_TDR} and the more optimistic CMS
simulations\cite{CMS_allHiggs}.

For the computation of the various branching ratios of the Higgs
and its couplings we rely on {\tt HDECAY}\cite{HDECAY} in which
the Higgs masses are determined following the two-loop
renormalisation group approach\cite{CarenaWagner_Higgs_Approx1}.

Since appreciable effects in the Higgs search occur for relatively
light spectra, this means that the light particles should also be
produced at an appreciable rate at the LHC even though they are
electroweak processes. We have calculated, at leading order, all
associated chargino and neutralino cross sections.
\beqn
pp \ra  \chi^\pm_i \chi^0_j   \;\;\;\;\; i=1,2\;;\;j=1,2,3,4
\eeqn

Neutralino pair production\footnote{K-factors for these processes
have been computed in \cite{Zerwaschichi-Kfactors}.} $ pp \ra
\chi^0_j \chi^0_k$ is much smaller with the heavy squark masses
that we assume. These processes have been calculated with the help
of {\tt CompHEP}\cite{comphep}. For the structure function we use
CTEQ4M at a scale, $Q^2=\hat{s}/4$.

It is also possible for the heaviest of these neutralinos to
cascade into the lighter ones and the lightest Higgs. We have
therefore calculated all branching ratios for all the charginos
and neutralinos. In principle other means of neutralino/chargino
production are possible through cascade decays of heavy squarks,
if these are not too heavy to be  produced at the LHC.

\section{Gauginos masses unified \`a la GUT}
\subsection{The available parameter space}
\begin{figure*}[hbtp]
\begin{center}
\includegraphics[width=10cm,height=8cm]{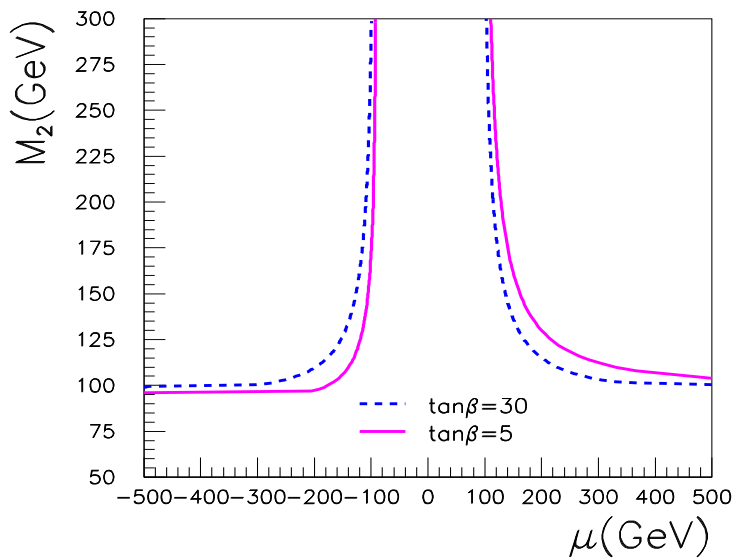}
\caption{\label{lim-unif-nodeg-189}{\em LEP2 excluded region
(inside the branch) in the $M_2-\mu$ parameter space for $\tgb=5$
and $\tgb=30$ from $m_{\chi_1^+}\geq 94.5GeV$.}}
\end{center}
\end{figure*}

In the case of no-degeneracy of the lightest chargino with  the
sneutrino, the constraint comes essentially from the chargino
cross section. With heavy sleptons, neutralino production does not
constrain the parameter space any further. Therefore the $\tgb$
independent limit Eq.~\ref{mchi90} applies. All these limits map
into the $M_2-\mu$ parameter space for a specific $\tgb$. The
available parameter space for $\tgb=5,30$ is shown in
Fig.~\ref{lim-unif-nodeg-189}

The absolute limit on the lightest neutralino for $\tgb=5$ turns
out to be:
\beqn\label{mneu90}
\mneuto \ge 47.5 GeV ( \tgb=5).
\eeqn
Therefore  in  the non degenerate case there is a very small
window for the Higgs to decay into neutralinos. For the lower
limit on the neutralino mass the reduction factor brought about by
the $\beta^3$ P-wave factor in Eq.~\ref{gamhtolsp} factor amounts
to about $.1$, for $m_h=109$GeV.

\subsection{The $A_t$ \tgbt dependence}
\begin{figure*}[htbp]
\begin{center}
\mbox{
\includegraphics[width=8cm,height=8cm]{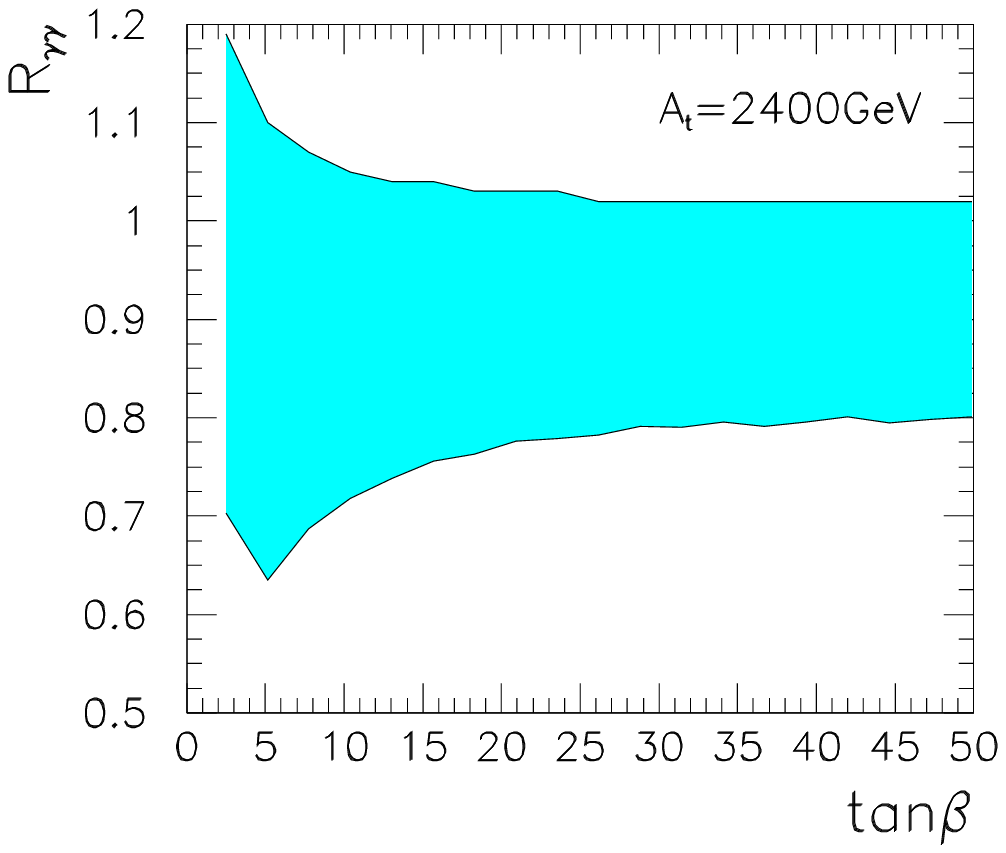}
\includegraphics[width=8cm,height=8cm]{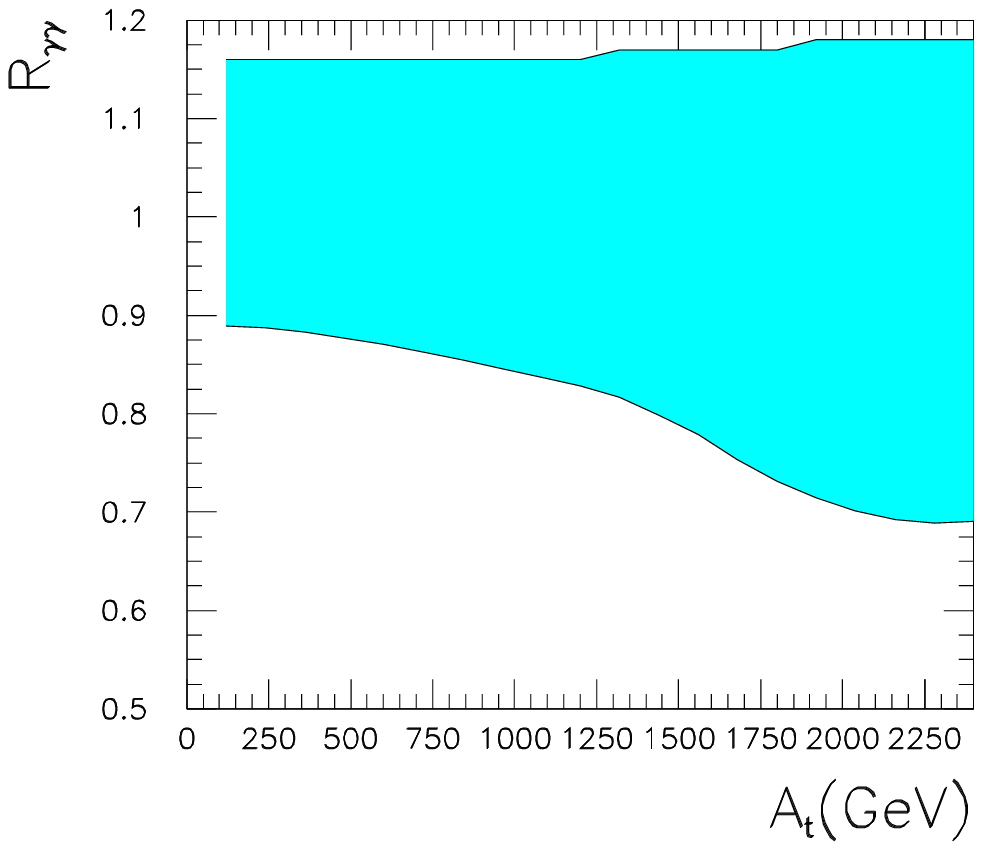}}
\caption{\label{tanxat2400}{\em $R_{\gamma\gamma}$ (shaded area)
as a function of a) $\tgb$ for $A_t=2.4$~TeV, b)$A_t$ for
$\tgb=2.5-50$ .}}
\end{center}
\end{figure*}

The above mass of the Higgs for $\tgb=5$ corresponds to a mixing
angle in the stop sector $A_t=0$. Obviously to maximise the effect
of the neutralinos through the opening up of the Higgs decay into
neutralino one should increase the mass of the Higgs. We have
already taken $M_A=M_S=m_{\tilde{t}}=\mgluino=1TeV$. We can
therefore increase $A_t$ and \tgbt. However increasing \tgbt also
increases the neutralino masses and reduces the $h\neuto \neuto$
couplings as we discussed earlier. Scanning over $\mu (>0)$, $M_2$
and \tgbt we show, Fig.~\ref{tanxat2400},  the extremal  variation
of the $R_{\gamma \gamma}$ as a function of \tgbt for maximal
mixing and taking the available constraints into account. We see
that the maximum drop is for $\tgb \sim 5$.  Below this value of
\tgbt the Higgs mass is small compared to the neutralino
threshold, while above this value the LSP gets heavier ``quicker"
than does the Higgs. Moreover the Higgs coupling to the LSP gets
weaker as $\tgb$ increases. On the other hand the increase
$R_{\gamma \gamma}>1$ grows with smaller \tgbt, but this is mainly
due to the loop effects of the charginos.
Also, as expected, the variation with $A_t$ affects essentially
the maximal reduction curve.

This said, let us however not forget that especially in the
two-photon signal at the LHC the significance increases with
increasing Higgs mass.  One can already conclude on the basis of
Fig.~\ref{tanxat2400} and our benchmark  $R_{\gamma \gamma}>.6$,
that critical regions are for moderate $\tgb$,  $\tgb \sim 5$, and
maximal stop mixing.

\def\rgg{R_{\gamma \gamma}}
\def\rggt{$R_{\gamma \gamma}$}

\subsection{The case with $A_t=0$ and $\tgb=5$}
\begin{figure*}[hbtp]
\begin{center}
\includegraphics[width=15cm,height=15cm]{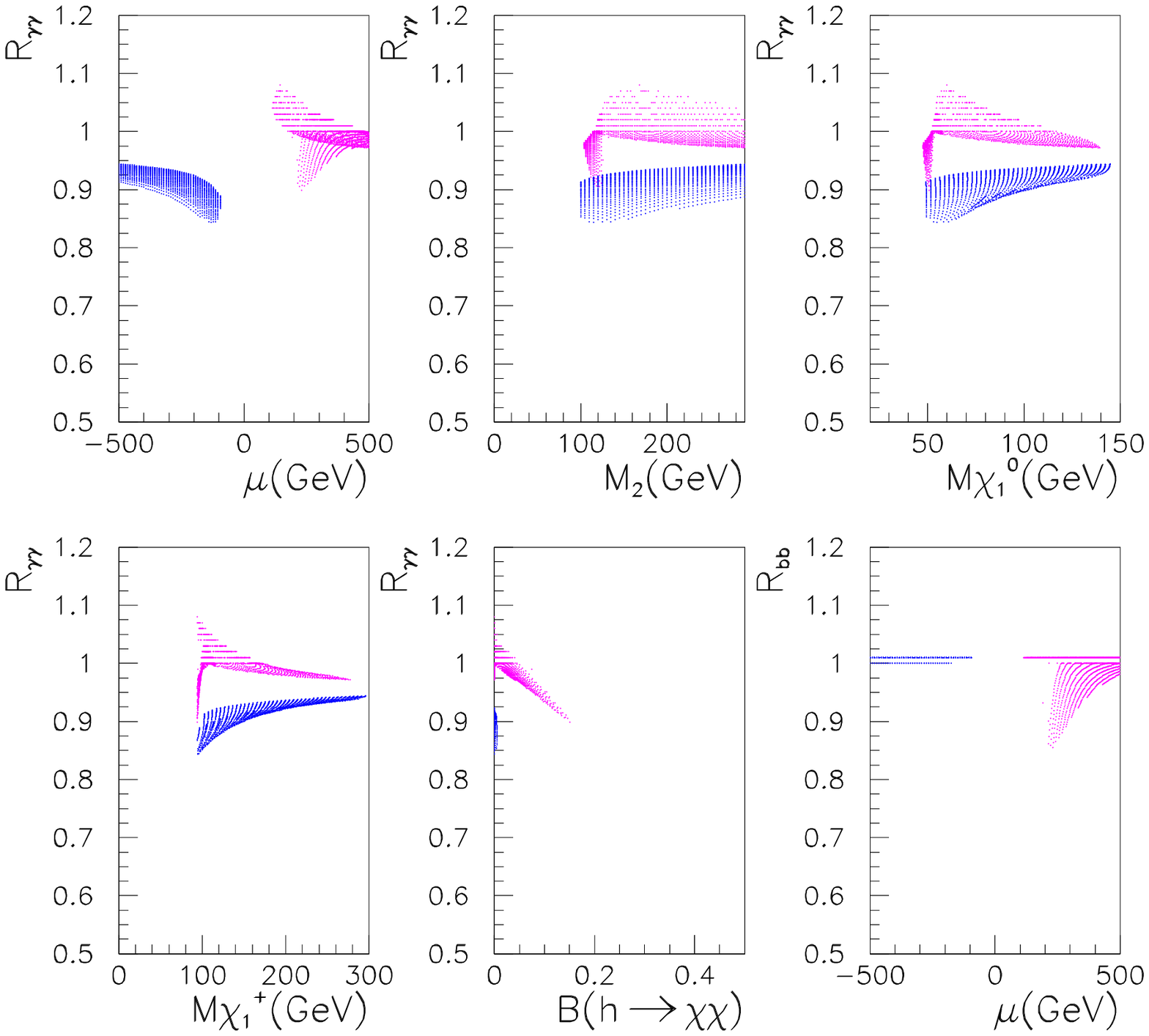}
\caption{\label{tan5at0}{\em Variation of $R_{\gamma\gamma}$ and
$R_{b \bar b}$ {\it vs} $\mu$, $M_2$, $m_{\chi_1^+}$, $\mneuto$
and $Br(h \ra \neuto \neuto)$. Also shown, last frame, the
variation of $R_{b \bar b}$ with $\mu$. All plots are for
$\tgb=5$, $M_2=50-300$GeV, $\mu=100-500$GeV and $A_t=0$. }}
\end{center}
\end{figure*}

We now go into more detail and choose $\tgb=5$ in the case of no
mixing. The results are summarised in Fig.~\ref{tan5at0}.

First of all note that in this scenario the ratio $R_{\gamma
\gamma}$ can vary at most by $15\%$ and that this can lead to
either a slight increase or a slight decrease. Contrary to what we
will see for other scenarios, the largest drop occurs for {\em
negative} values of $\mu$ and is due to the contribution of the
light charginos in the two-photon width (see also the dependence
with $\mchargo$ and $M_2$). The sign of $\mu$ is also that of the
interference between the dominant $W$ loop and the chargino loop
contribution. A decrease for positive $\mu$ is strongly correlated
with the opening up of the little window for $h\ra \neuto \neuto$.
The latter channel leads to a branching ratio which is at most
some $20\%$. When this occurs (only for positive $\mu$) it will
affect also the branching into $b \bar b$ and thus the channel $t
\bar t h \ra t\bar t b \bar b$. However with our benchmark for
observability of the Higgs in this channel, $R_{b \bar b}>.7$, the
Higgs should still be observed in this channel.

At this stage one can conclude that the effect of light
charginos/neutralinos, especially in view of the theoretical
uncertainty (higher order QCD corrections) in predicting the
signal, is very modest. Furthermore the small window for Higgs
decaying into LSP will be almost closed, at least at $\tgb=5$,
with an increase of a few GeV  on the lower limit on charginos.

\subsection{The case with maximal $A_t$ and $\tgb=5$}
\begin{figure*}[htbp]
\begin{center}
\includegraphics[width=14cm,height=15cm]{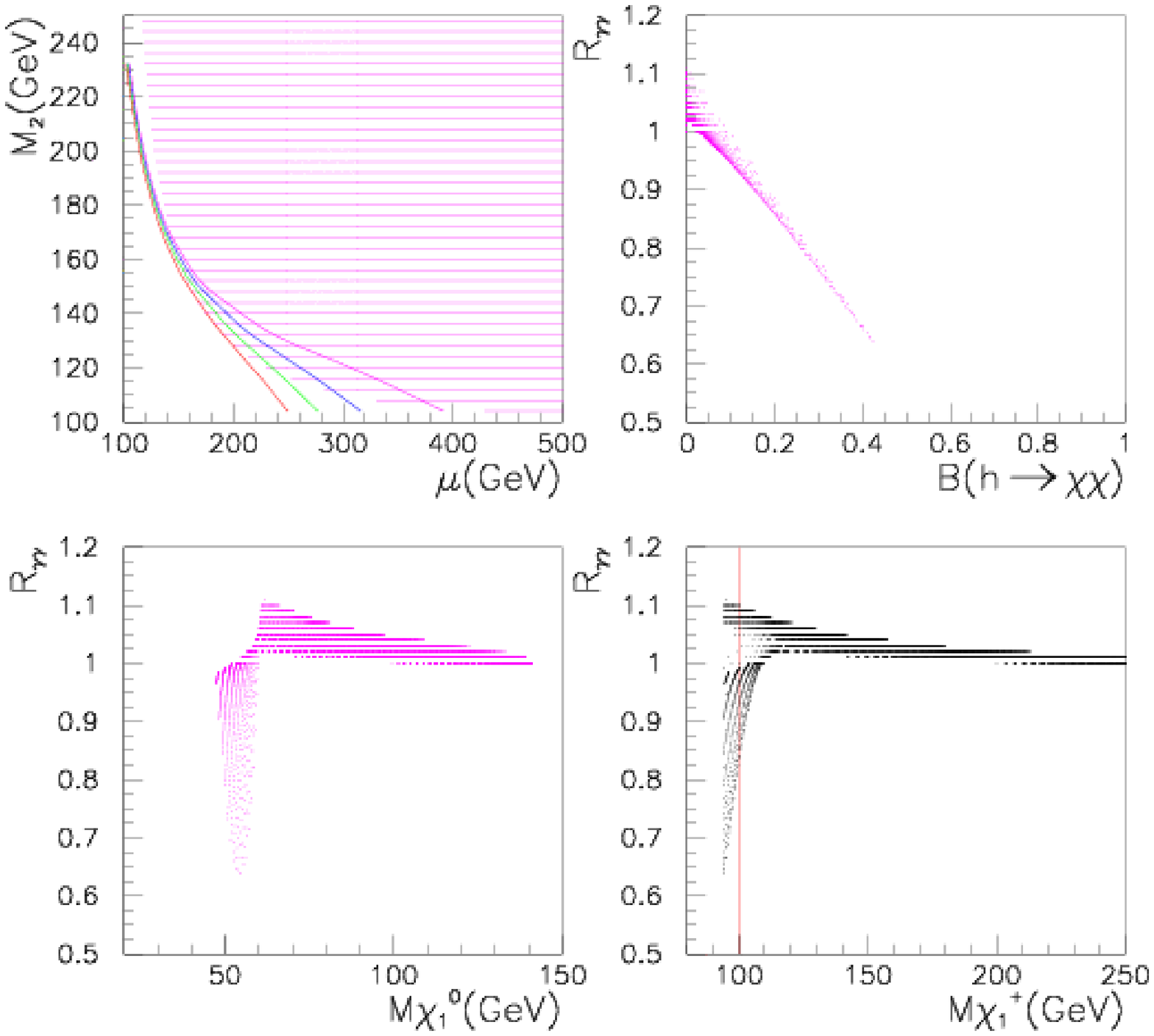}
\caption{\label{tan5at2400} {\em a) Contour plot of $Br(h\ra
\neuto \neuto)=0.1,0.2,0.3,0.4$ (from right to left respectively)
in the $M_2-\mu$ plane. The shaded area is the allowed region.b)
Correlation between \rggt and $Br(h\ra \neuto \neuto)$  c)
Variation of $R_{\gamma\gamma}$ with  the mass of the LSP
$M_{\chi_1^0}$ and d) mass of the chargino $M_{\chi_1^+}$. The
vertical line corresponds to $\mchargo=100$GeV. All plots are for
$\tgb=5$, $M_2=50-300$GeV, $\mu=100-500$GeV and $A_t=2.4$TeV \/.}}
\end{center}
\end{figure*}

Increasing the mass of the Higgs through as large $A_t$ as
possible for the same value of \tgbt changes the picture quite
substantially. With our implementation of the corrections to the
Higgs mass the increase is about 10GeV and leaves enough room for
$h \ra \neuto \neuto$ in the small $\mu-M_2$ region. In this case
the two-photon rate and the $h\ra \neuto \neuto$ branching ratios
are well correlated as shown (Fig.~\ref{tan5at2400}a), the result
of a scan over the parameters $M_2=50-300$~GeV, $\mu=100-500$~GeV
for $\tgb=5$ in the maximal mixing case, $A_t=2.4$~TeV. A scan
over a wider range, $M_2\leq 2$~TeV and $|\mu|\leq 1$~TeV,  was
also performed. The points for larger values of $M_2-\mu$ all
cluster around $\rgg \approx 1$ allowing for only a few percent
fluctuations. The Higgs branching ratio into neutralinos can reach
as much as $40\%$, leading to a reduction of \rggt and $R_{b \bar
b}$ of about $60\%$. This means that there might be problems with
Higgs detection especially in the $t\bar t h$ channel. The contour
plots of constant $h\to \neuto \neuto$ in the $M_2-\mu$ plane are
displayed in Fig.~\ref{tan5at2400}b). It is only in a small region
$M_2 \leq 160$~GeV and $\mu\leq 400$~GeV that $h\ra \neuto \neuto$
exceeds 10\%.

As the results presented here depend critically on the minimum
allowed value for the mass of the lightest chargino and
neutralino, see Fig.~\ref{tan5at2400}c-d), it is   interesting to
enquire about the consequence of an improved lower limit of the
chargino masses in the last runs of LEP2. We have therefore
imposed the constraint $\mchargo \geq 100$~GeV. As the maximum
reduction occurs for the lightest allowed value for the chargino
mass, an increase of just a few GeV's has a drastic effect. The
reduction in \rggt$\;$ is no longer more than $80\%$. In
conclusion, the effect of gauginos/higgsinos on the crucial
branching ratio of the Higgs, when one assumes the unification
condition and no degeneracy,  will only be marginal at the LHC if
LEP2 does not observe any  charginos or neutralinos before the end
of its final run.

\subsection{Associated chargino and neutralino production at the
LHC}
\begin{figure*}[hbtp]
\begin{center}
\includegraphics[width=10cm,height=10cm]{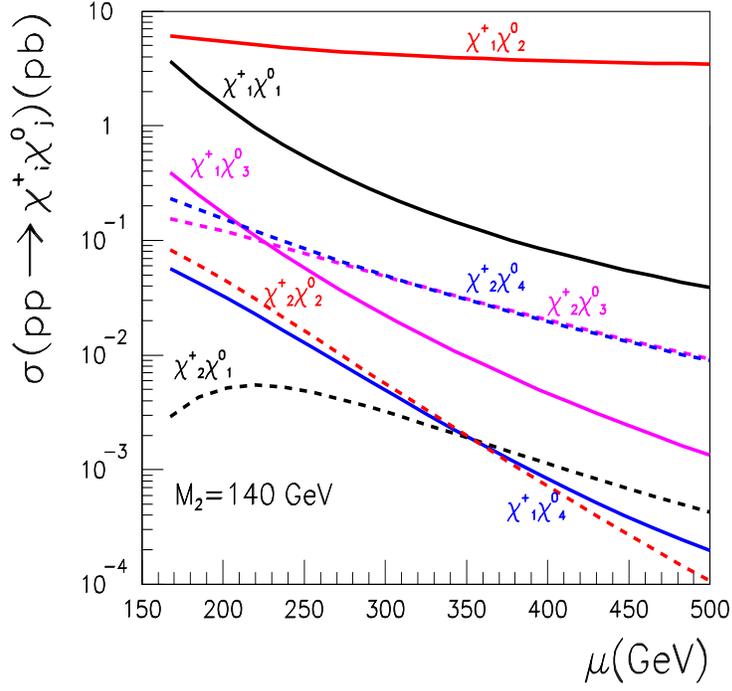}
\includegraphics[width=10cm,height=10cm]{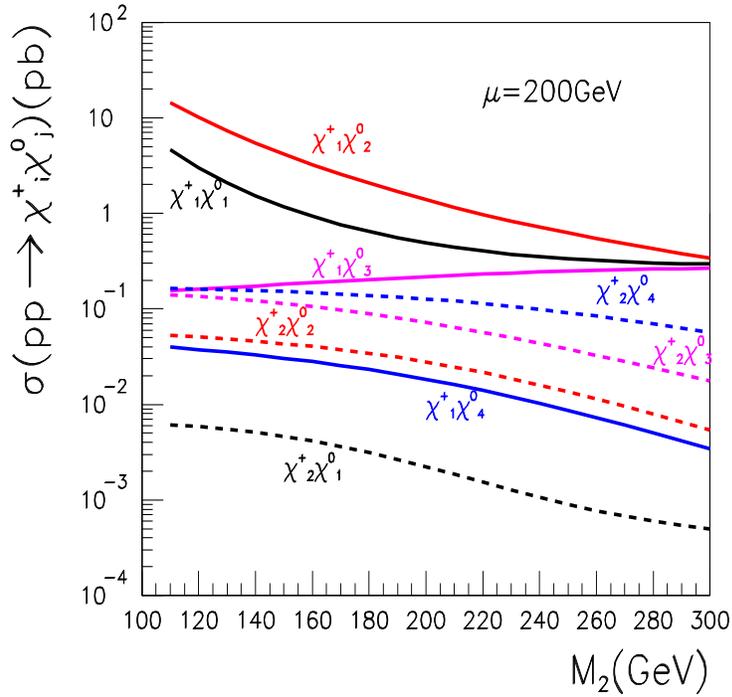}
\caption{\label{chiprod-unif}{\em Associated Production of
chargino and neutralino at the LHC at LO  a) as a function of
$\mu$ for $M_2=140$GeV. b) as a function of $M_2$ for
$\mu=200$GeV. In the case of gaugino mass unification without
degeneracy with the sneutrinos, the LEP limit means that for
$M_2=140$GeV, $\mu>176$GeV and for $\mu=200$GeV, $M_2>130$GeV. In
the case of degeneracy with the sneutrinos all points in the
figure are valid \/.}}
\end{center}
\end{figure*}
In our previous study of the effects of light
stops\cite{nous_Rggstophiggs_lhc} on the Higgs search at the LHC,
reduction in the usual two photon signals was due essentially to a
drop in the main production mechanism through gluons and occurred
when the stops developed strong couplings to the Higgs. When this
occurs, as a lever, one has large production of stops as well as
associated stop Higgs production, thus recovering a new mechanism
for Higgs. In the present case uncovering a new effective Higgs
production mechanism will be more complicated. First the effects
are due to weakly interacting particles whose cross sections at
the LHC are smaller than those for stops. Also since the largest
drops are when the branching ratio of the Higgs  into invisible is
appreciable, this means that even if one triggers Higgs production
through charginos and neutralinos , the reconstruction of the
Higgs will be more difficult. Nevertheless one should enquire how
large any  additional production mechanism, if any, can get. In
the present scenario with a common gaugino mass at the GUT scale
and no (accidental) degeneracy between the chargino and the
sneutrinos, $R_{\gamma \gamma}$ (and $R_{b\bar b}$) being at worst
$.6$ (for maximal mixing), the Higgs should be discovered in the
usual channels. Moreover  since the $Br(h\ra b\bar b)$ does not
drop below about $.6$, we could use this signature in the cascade
decay of the heavier neutralinos and charginos into Higgs.

\begin{figure*}[hbtp]
\begin{center}
\includegraphics[width=15cm,height=7cm]{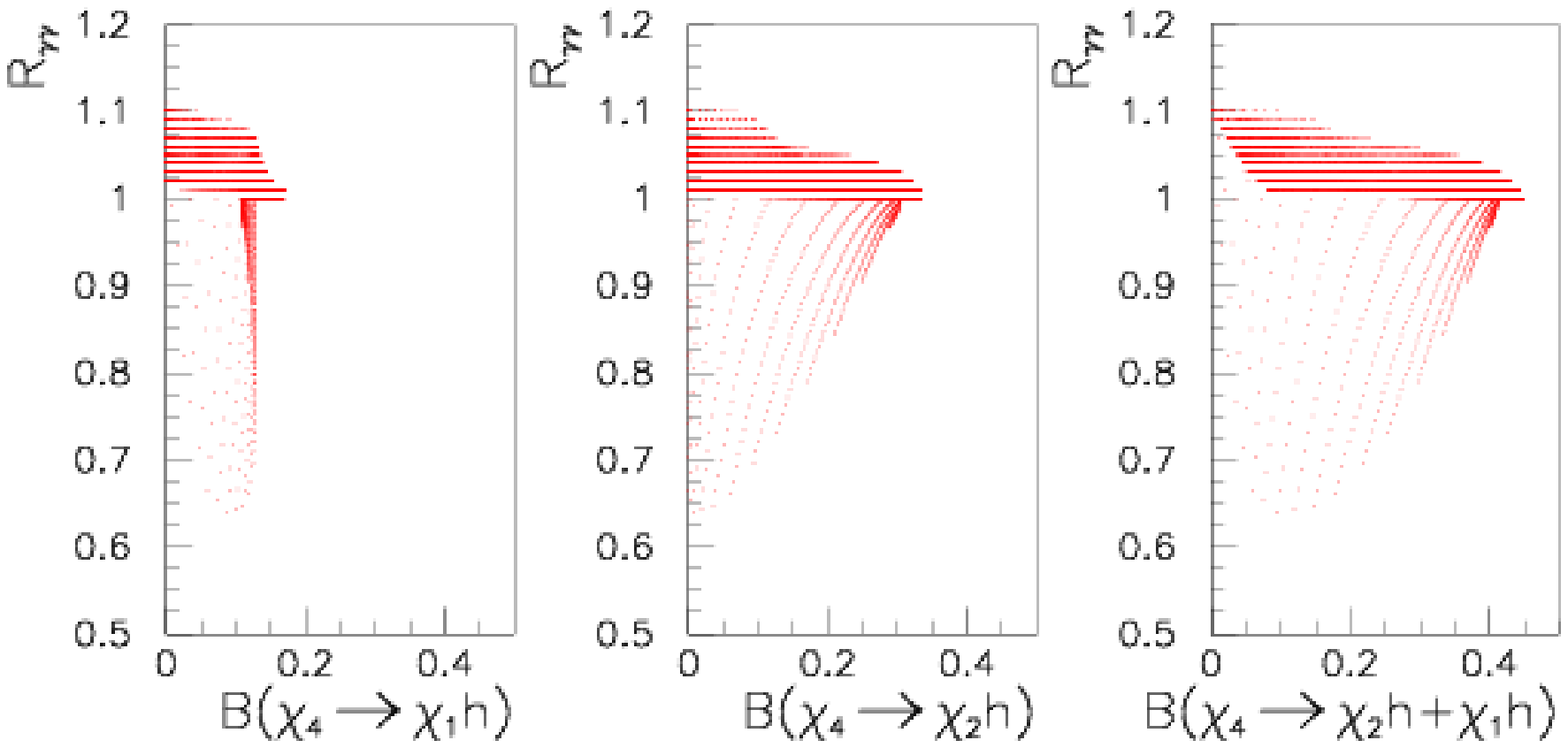}
\caption{\label{tan5br}{\em Branching ratio of $\neutf$ into $h$
for $\tgb=5$, $A_t=2400$GeV. The scan over $M_2$ and $\mu$ is as
in Fig.~\ref{tan5at0}\/.}}
\end{center}
\end{figure*}

Since the  reduction in the usual inclusive two photon channel
always occurs in the small $(M_2,\mu)$ region, all gauginos are
relatively light and therefore have reasonable production rates.
In fact as Fig.~\ref{chiprod-unif} shows, the rates are more than
reasonable in the parameter space that leads to the largest drops.
For instance, with $M_2=140$GeV, the cross section $\neutt
\chi_1^+$ is about $6$pb and is mildly dependent on $\mu$, while
production of $\neutf \chi_2^+$,  is some $100$fb (with $\mneutf
\sim 250$GeV) when $R_{\gamma \gamma}=.6$, and decreases quickly
with increasing $\mu$ (where however $R_{\gamma \gamma}$
increases). With the first process, considering the rather large
cross section, it should be possible through measurements of the
masses and some of the signatures of $\neutt$ and $\chi_2^+$ to
get some information on the parameters of the neutralinos and
charginos\footnote{See for instance \cite{nojiri-lhc}.}, we would
then know that one might have some difficulty with the Higgs
signal in the inclusive channel. As for the latter process, it has
more chance to trigger light Higgs than the former. Since in our
scenario there isn't enough phase space for $\neutt \ra \neuto h$.
The following modes are potentially interesting: $\chi_4^0 \ra
\chi_{1,2}^0 h$ and $\chi_2^+ \ra \chi_1^+ h$. For the former one
obtains as much as $25\%$ branching ratio for $\neutf \ra h +{\rm
anything}$ when \rggt is lowest , see Fig~.~\ref{tan5br}. Much
higher branching are of course possible, but they occur for higher
values of $\mneutf$ where there is no danger for Higgs discovery
in the usual modes. Less effective and not always open is the mode
$\chi_3^0 \ra \chi^0_{1,2} h$ where the branching never exceeds a
few per-cent.

\begin{figure*}[hbt]
\begin{center}
\includegraphics[width=15cm,height=10cm]{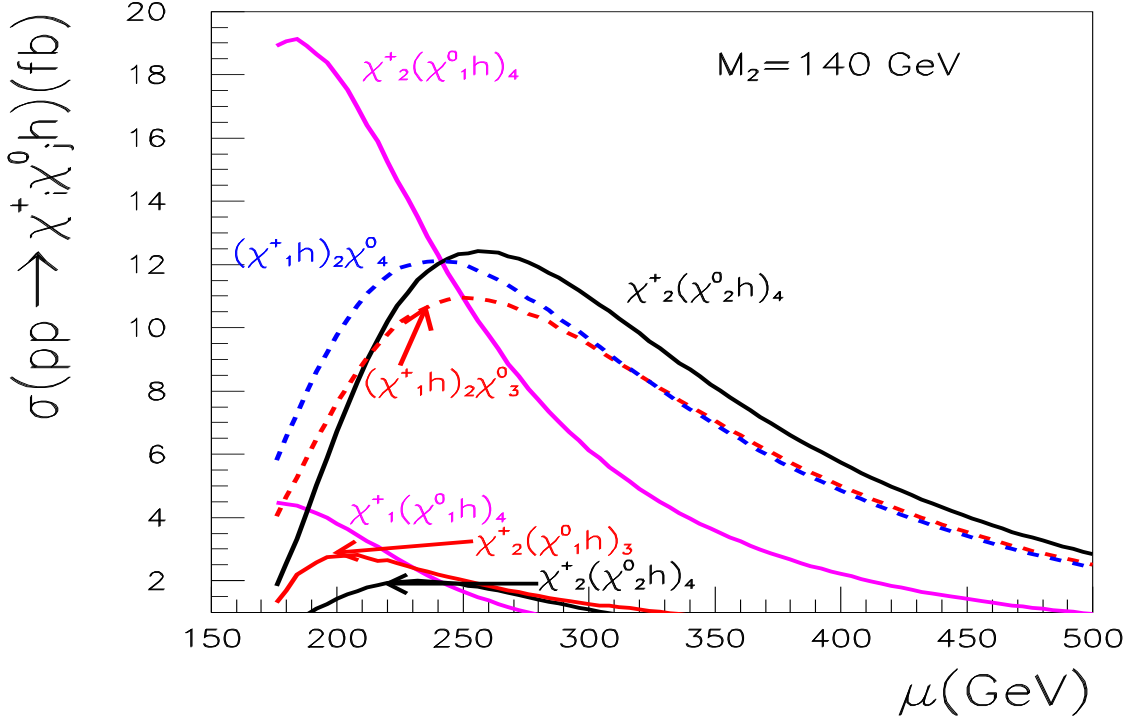}
\caption{\label{folding-uni-nondeg}{\em Higgs yield through
charginos and neutralinos decays as a function of $\mu$.
$M_2=140$GeV and $\tgb=5$ and maximal mixing. The subscript for
the parentheses $(\;\;)_j$ indicates the parent neutralino or
chargino\/.}}
\end{center}
\end{figure*}

We are now in a position of folding the different branching ratios
for the heavier neutralinos and  chargino into Higgs (h) with the
corresponding cross sections to obtain the yield of Higgs in these
channels. As advertised, for the parameters of interest, we see
from Fig.~\ref{folding-uni-nondeg} that the largest cross sections
originate from the decays of the heaviest neutralino $\neutf$
while the chargino helps also. Still, the yield is quite modest,
about $20$fb. It rests to see whether a full simulation with a
reduced branching ratio of $h$ into b's can dig out the Higgs
signal from such cascade decays. We should make another remark. In
\cite{gluinoscascadetoh}, where $\neutt \ra \neuto h$ and $h\ra b
\bar b$ is advocated, the neutralinos themselves are produced
through cascade decays of gluinos and squarks which can have large
cross sections. In our case we have taken these to be as heavy as
$1$TeV and thus their cross section is rather modest. For instance
gluino pair production at the LHC with this mass is about $.2$pb.
However, without much effect on the decoupling scenario we have
assumed, if we had taken $\mgluino=500$GeV, which by the way
corresponds to a situation where the gaugino mass unification
extends also to $M_3$, the gluino cross section jumps to about
$20$pb. So many gluinos could, therefore, through cascade decays
provide an additional source of Higgs.

\section{Gauginos masses unified \`a la GUT degenerate with sleptons }
In the so-called sneutrino-degenerate case where charginos can be
as low as $70$GeV, the absolute lower limit on the neutralino LSP
mass:
\beqn
\label{limit70}
 m_{\chi_0} \geq 34.5GeV ( \tgb=5).
\eeqn
This lower bound rises by roughly 1GeV for $\tgb=2.5$ and never
goes below $34$GeV for larger values of $\tgb$. We will only study
the case with $A_t$ maximum.

\subsection{Results}
\begin{figure*}[htbp]
\begin{center}
\includegraphics[width=14cm,height=15cm]{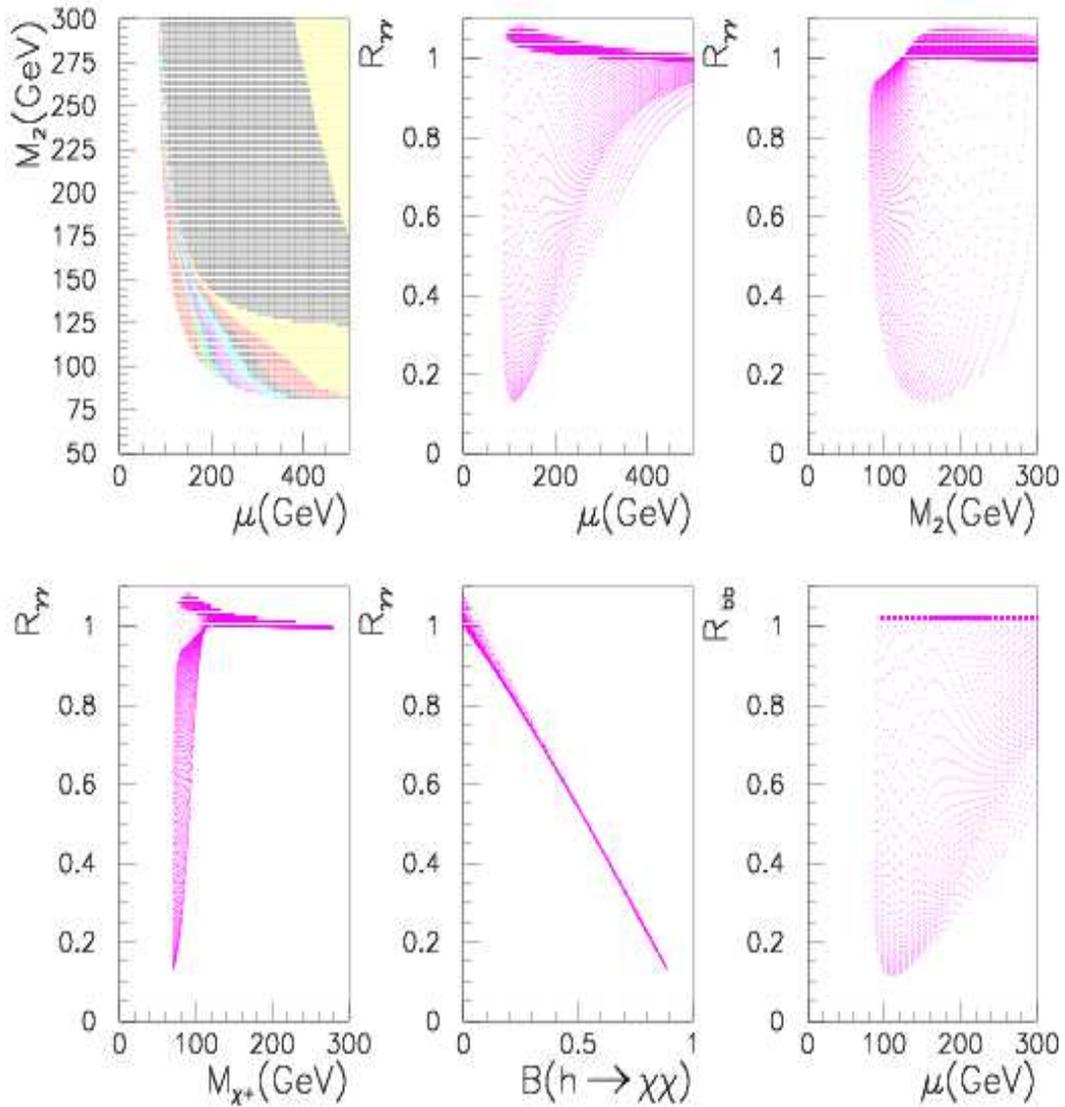}
\caption{ \label{resultsdeg} {\em Results in the degenerate
scenario with  $\tgb=5$ and maximal $A_t$.  In all plots the scans
are over $M_2=50-300$GeV, $\mu=100-500$GeV. From left to right and
top to bottom (a) Density plot for \rggt  in the allowed $M_2-\mu$
plane. The different shadings correspond to $ .3
<\rgg <1.1$ from left to right. (b) Variation of \rggt with $\mu$
(c) with $M_2$ (d) with the mass of the chargino $M_{\chi_1^+}$.
(e) Correlation between \rggt and the branching of $h$ into LSP.
(f) Variation of $R_{b \bar b}$ with $\mu$\/.}}
\end{center}
\end{figure*}
\begin{figure*}[hbtp]
\begin{center}
\includegraphics[width=15cm,height=10cm]{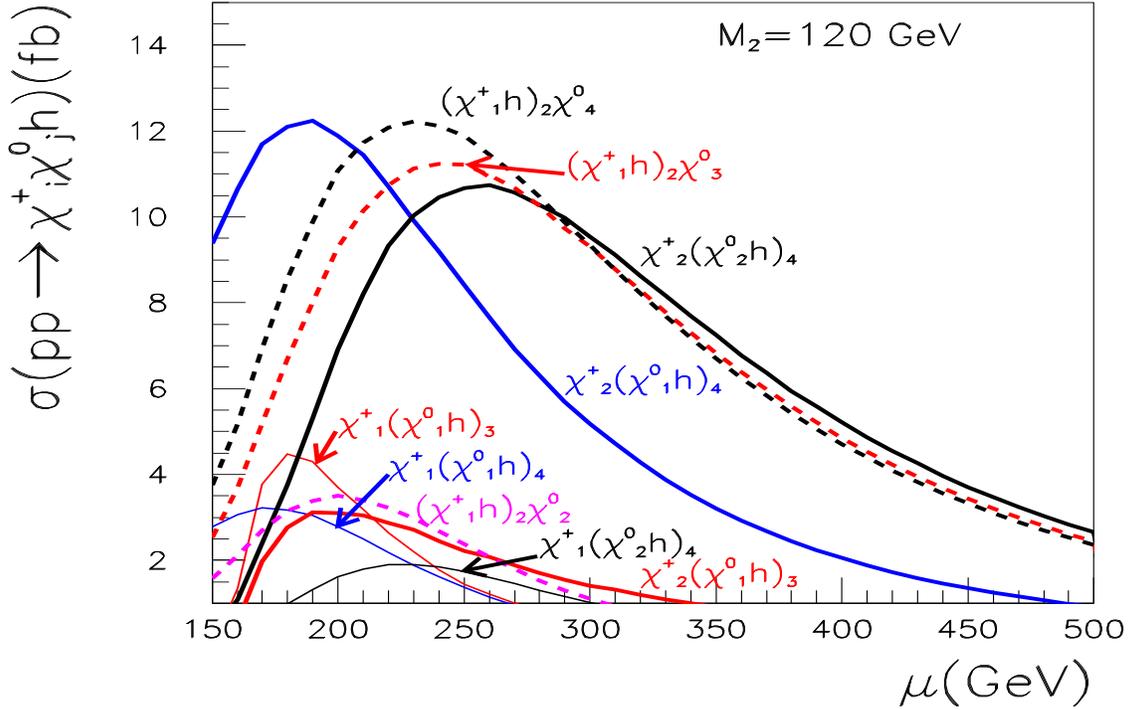}
\caption{\label{folding-uni-deg}{\em As in
Fig.~\ref{folding-uni-nondeg} but in the case of a chargino
degenerate with the sneutrino and $M_2=120$GeV\/.}}
\end{center}
\end{figure*}

Relaxing the chargino mass by some $20$GeV has quite impressive
effects that result in dramatic drops, see Fig.~\ref{resultsdeg}.
The branching fraction into invisibles can be as large as $90\%$.
For these situations clearly the Higgs would be difficult to hunt
at the LHC in both the two-photon  and (associated) $b \bar b$
channels. As seen for \rggt {\it vs} $\mchargo$, there is an
immediate fall for $\mchargo<100$GeV. But then this should be
compensated by the production of plenty of charginos and sleptons
while some of the heavier neutralinos and chargino should still be
visible. As indicated by Fig.~\ref{chiprod-unif}, in this
situation all charginos and all neutralinos will be produced with
cross sections exceeding $100$fb. $\chi^+_1$ has a cross section
in excess of $10$pb!. These processes can trigger Higgs
production.  However, because of the decays into light sleptons
the rates are modest as seen in Fig.~\ref{folding-uni-deg}. In
fact, the largest rates occur when the Higgs has the largest
branching into invisible. These modes will probably not help much.

\section{Relaxing the gaugino mass unification}
As we have seen, having light neutralinos can very much jeopardise
the Higgs discovery at the LHC. However in the canonical model
with $M_1 \simeq M_2/2$ and no pathological degeneracy the effect
is never a threat. Basically this is because the almost model
independent limit on the chargino translates into values of $M_2$
(hence $M_1$) and $\mu$ large enough that the neutralinos are not
so light that they contribute significantly a large invisible
Higgs width. On the other hand if $M_1$ were made much smaller
than $M_2$, one could make $\mneuto$ small enough without running
into conflict with the chargino mass limits. The LSP could then be
very light and almost bino. To make it couple to the Higgs though
one still needs some higgsino component and thus $\mu$ should not
be too large. Largest couplings will be for smallest values of
$\mu$ which are however, again, constrained by the chargino mass
limit for instance. To investigate such scenarios we  have studied
the case with
\beqn
\label{nounicdt} M_1=r \; M_2 \;\;\; {\rm with} \;\;\; r=0.1
\eeqn
and have limited ourselves to the case with $\tgb=5$.

Models with $r>1$ would not affect the Higgs phenomenology at the
LHC, since their lightest neutralino should be of the order of the
lightest chargino. LEP data already excludes such a neutralino to
contribute to the invisible width of the Higgs and therefore the
situation is much more favourable to what we have just studied
assuming the usual GUT relation.

It is important to stress that the kind of models we investigate
in this section are quite plausible. The GUT-scale relation which
equates all the gaugino masses at high scale need not be valid in
a more general scheme of SUSY breaking. In fact even within SUGRA
this relation need not necessarily hold since it requires the
kinetic terms for the gauge superfields to be the most simple and
minimal possible (diagonal and equal). One can easily arrange for
a departure from equality by allowing for more general forms for
the kinetic terms\cite{nmSUGRA}. In superstring models, although
dilaton dominated manifestations lead to universal gaugino masses,
moduli-dominated or a mixture of moduli and dilaton fields lead
also to non universality of the gaugino
masses\cite{nonuiniversal-strings} and may or may not
(multi-modulii\cite{multimoduli}) lead to universal scalar masses.
The recent so-called anomaly-mediated SUSY breaking
mechanisms\cite{non-universal-anomaly} are also characterised by
non-universal gaugino masses, though most models in the literature
lead rather to $r>1$ which is of no concern for the Higgs search.

With $r=1/10$ the main feature is that the neutralino mass
spectrum is quite different. Most importantly LSP have masses in
the range $\sim 10-20$GeV for the cases of interest. Since there
is plenty of phase space for the decay of the lightest Higgs into
such neutralinos we will only consider $A_t=0$ for the stop
mixing.

\subsection{The available parameter space}
\begin{figure*}[hbtp]
\begin{center}
\includegraphics[width=12cm,height=8cm]{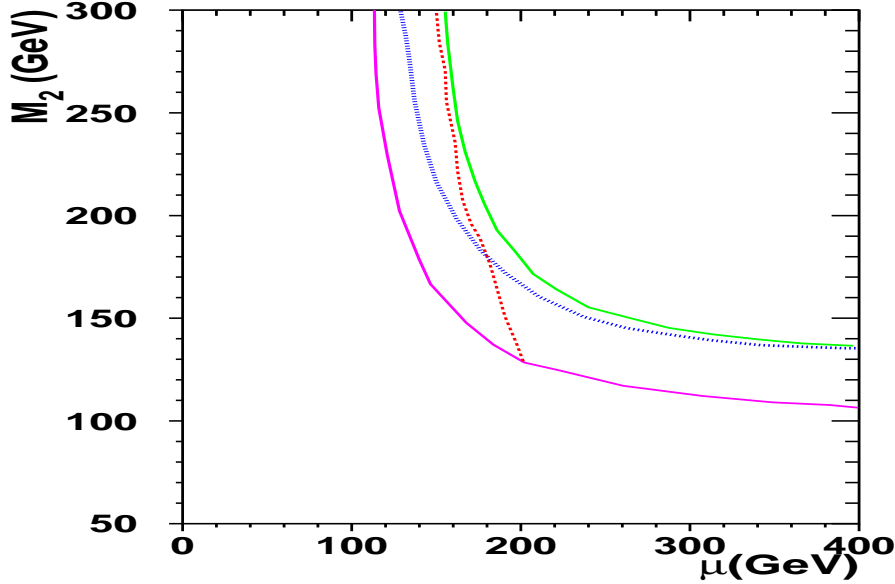}
\caption{\label{lim-nounif189}{\em LEP2 allowed region in the
$M_2-\mu$ parameter space in the case $M_1=M_2/10$ for $\tgb=5$
and $\mu>0$. The inside curve corresponds to the case where all
sleptons are heavy ($1TeV$) and is derived essentially from the
chargino cross section. The dashed curve is for $\mser=100$GeV and
$\msel=1$TeV, the dotted curve is with  $\msel=150$GeV and
$\mser=1$TeV, the outer curve is for $\msel=150$GeV and
$\mser=100$GeV. The limits in the case of light sleptons are
derived from data on neutralino production, see text.}}
\end{center}
\end{figure*}
In the case of heavy sleptons we find that the $\mu-M_2$ allowed
parameter space is still determined from the chargino mass limit
through $\epem \ra \chi^+_1 \chi^-_1$ production. Neutralino pair
production $\neuto \neutt$ and $\neuto \neutth$, although
kinematically possible do not squeeze the parameter space further.
The contour plot, see Fig.~\ref{lim-nounif189}, is therefore
essentially the same as the one with the GUT relation. Since
cosmological arguments will drive us to consider light sleptons
masses, we show on the same figure, Fig.~\ref{lim-nounif189}, how
the $\mu-M_2$ parameter space is squeezed in this case. The
squeezing comes from limits on $\neuto \neutt$ and $\neuto
\neutth$ cross sections properly folded with branching ratios
where two-body and three-body decays involving the relatively
light sleptons play an important role. In fact, while light
sleptons generally enhance the neutralino cross sections, this
enhancement can be counterbalanced by the fact that a non
negligible branching ratio into invisible neutrinos can occur with
small enough left selectrons. In all cases the leptonic final
state signature can be enhanced at the expense of the hadronic
signature which usually have a better efficiency. To illustrate
this, we have considered three cases: i) $\mser=100$GeV with large
$\msel$, ii) $\msel=150$GeV with large $\mser$ iii)$\mser=100,
\msel=150$GeV. One sees that, with a very mild  $M_2$ dependence,
light right selectrons eliminate smallest $|\mu|$ values that are
otherwise still allowed by chargino searches. That $\ser$ do not
cut on $M_2$ values can be understood on the basis that they do
not have any $SU(2)$ charge. Since smallest values of $\mu$ are
the ones that enhance $h \ra \neuto \neuto$ these limits are
important. With light left selectrons  the gain with respect to
the chargino limit is appreciable and occurs across all $M_2$
values, more so for  the smallest $M_2$ values. When both left and
right selectrons are relatively light, one carves out an important
region, although this region does not cover all the available
neutralino phase space.
\subsection{Heavy sleptons}
\begin{figure*}[htbp]
\begin{center}
\includegraphics[width=16cm,height=18cm]{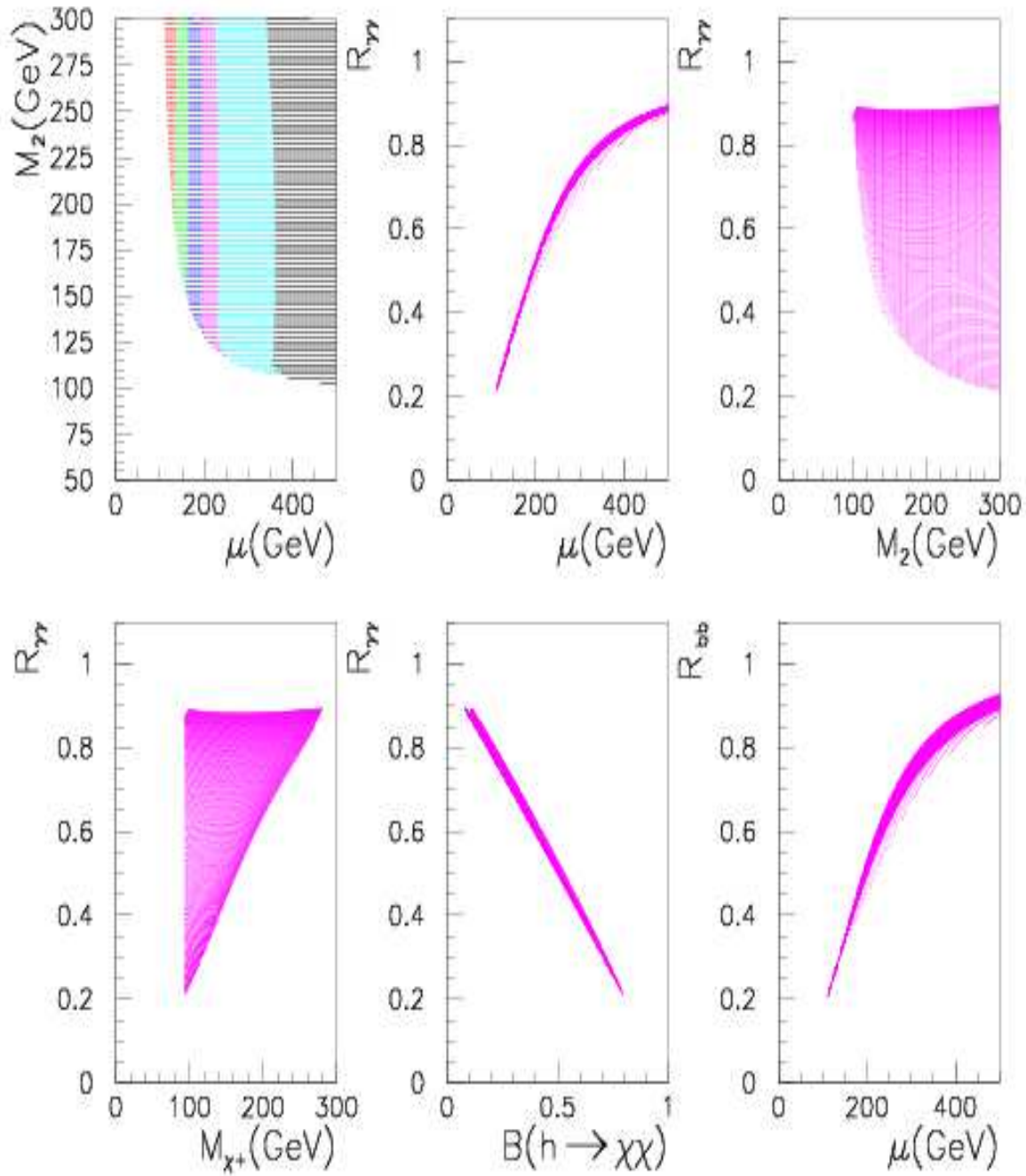}
\caption{\label{no-unif-heavysl} {\em Effects of neutralinos from
$M_1=M_2/10$ with $\tgb=5$ and $A_t=0$ with heavy selectrons. In
all the plots, scans are over $M_2=50-300$GeV, $\mu=100-500$GeV.
From left to right and top to bottom a) Density plot for \rggt  in
the allowed $M_2-\mu$ plane. The different shadings correspond to
$ .3 <\rgg <.4$ (left band) to $ .8 <\rgg <.9$ (right band). b)
Variation of \rggt with $\mu$ c) with $M_2$ d) with the mass of
the chargino $M_{\chi_1^+}$. e) Correlation between \rggt and the
branching into LSP. f) Variation of $R_{b \bar b}$ with $\mu$\/.}}
\end{center}
\end{figure*}

The main message is that there are some dangerous reductions in
the branching ratios of the Higgs both into photons and into $b
\bar b$ which can be only a $1/5$th of what they are in the \sm,
see Fig.~\ref{no-unif-heavysl} . These drops are due essentially
to a large branching ratio of the Higgs into invisibles. The most
dramatic reductions occur for chargino masses at the edge of the
LEP2 limits, however even for chargino masses as high as $200$GeV
the drop can reach $60\%$. In these configurations the lightest
chargino and $\neutt, \neutth$ have a large higgsino component.
This explains why, in the $M_2-\mu$ plane the decrease in the
ratios is strongly dependent on $\mu$.

\subsection{Cosmological constraint}
\begin{figure*}[htbp]
\begin{center}
\includegraphics[width=14cm,height=15cm]{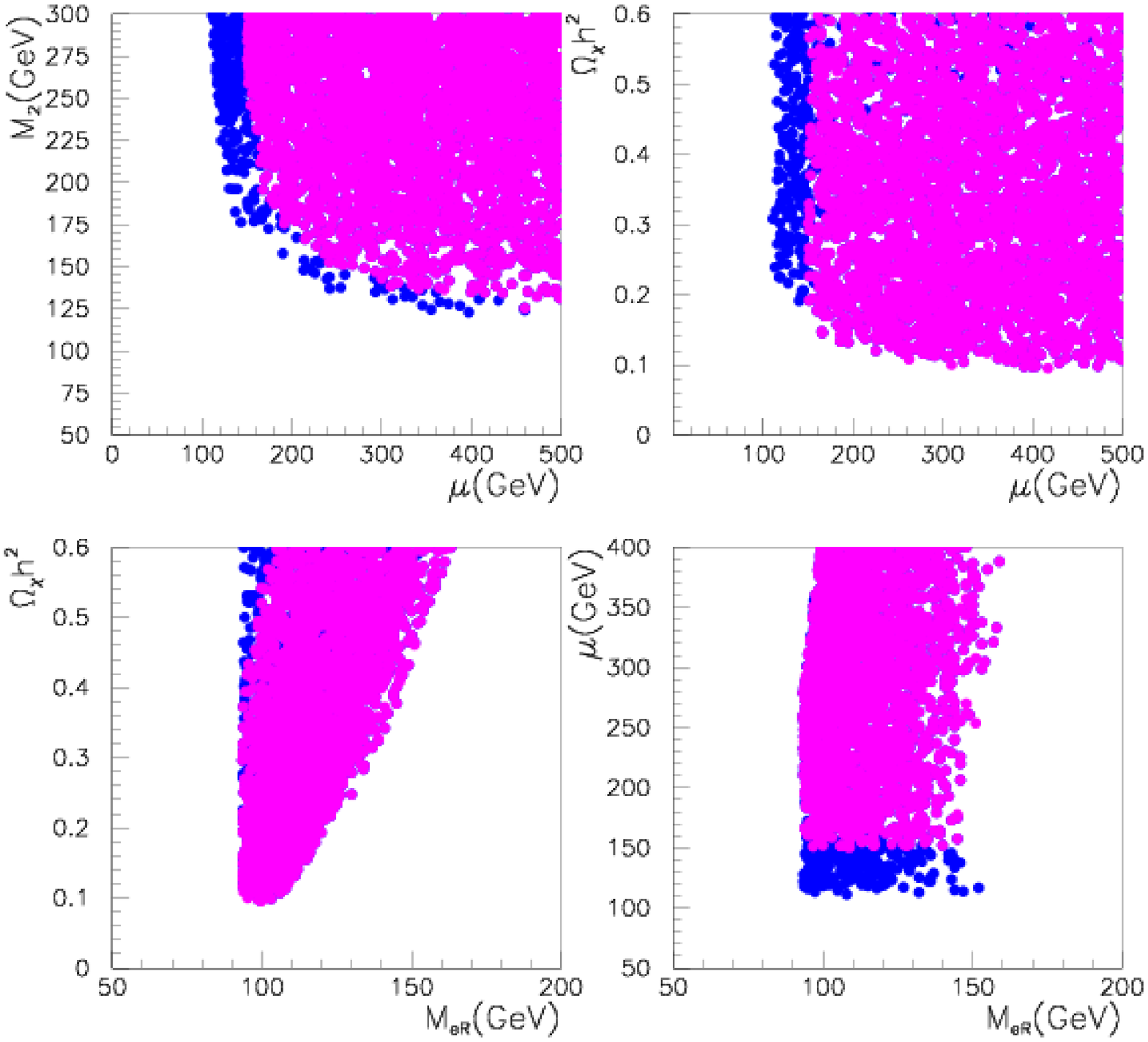}
\caption{\label{relic-nounif} {\em All points that pass the
constraint $\Omega_\chi h^2 <.6$ in the case $M_1=M_2/10$ with
$\tgb=5$ and $A_t=0$. In  these plots, scans are over
$M_2=50-300$GeV, $\mu=100-500$GeV and $m_0=50-1000$GeV. From left
to right and top to bottom a) Allowed region in the $M_2-\mu$
plane.  b) $\Omega_\chi h^2$ {\it vs} $\mu$ c) $\mser$ d) Scatter
plot of $\mu$ {\it vs} $\mser$. The grey areas represent the
allowed parameter space after including the constraints from
neutralino cross sections at $\sqrt{s}=189$GeV while the extra
dark area does not include the constraint on neutralino
production\/.}}
\end{center}
\end{figure*}
Considering these large reductions and the fact that the LSP is
very light, $10-20$GeV, we investigated whether the most dramatic
scenarios are not in conflict with a too large relic
density\footnote{Cosmological consequences of non-unified gaugino
masses have been investigated in \cite{m2m1-10-astro} but not from
the perspective followed in this paper.}. One knows that for a
very light LSP bino the annihilation cross section is dominated by
sfermions with largest hypercharge, that is right
sleptons\cite{relic-classic,astrosusy-review}. This calls for
light (right) sfermions. As a rule of thumb, with all sfermions
heavy but the three right sleptons, an approximate requirement is
\beqn
\label{approx-relic} m_{\tilde{l}_R}^2 < 10^{3} \sqrt{(\Omega_\chi
h^2)_{\rm max}} \times \mneuto.
\eeqn
with all masses expressed in GeV.

In our case the LSP is not a pure bino, the bino purity is around
$90\%$ for the worst case scenarios, otherwise it would not couple
to the Higgs. We have therefore relied on a full calculation. We
assumed  all squarks heavy and took a common mass for the SUSY
breaking sfermion mass terms of both left and right sleptons of
all three generations, $m_0$, defined at the GUT scale, thus
assuming unification for the scalar masses. As for the gaugino
masses to obtain $M_1=M_2/10$ at the electroweak scale one needs
$\bar{M}_1 \simeq \bar{M}_2/5$ at the GUT scale. $\bar{M}_2$ is
the $SU(2)$ gaugino mass at the GUT scale which again relates to
$M_2$ at the electroweak scale as  $M_2 \sim 0.825 \bar {M}_2$.
This scheme  leads to almost no running of the right slepton mass,
since the contribution from the running is of order $M_1^2$, while
left sleptons have an added $M_2^2$ contribution and would then be
``much heavier". Indeed neglecting Yukawa couplings one may write
\beqn
\label{m0runing} \mser^2&=&\bar{m}_0^2\;+\;
0.006\bar{M}_2^2\;-\;\sww D_z \nonumber
\\ \msel^2&=&\bar{m}_0^2\;+ 0.48\bar{M}_2^2\; -\;(.5-\sww)D_z \nonumber
\\ \msnue^2&=&\bar{m}_0^2\;+ 0.48\bar{M}_2^2\;+\;D_z/2  \;\;\;\;\;\;
\eeqn

\noi Note in passing that Eq.~\ref{m0runing} can be extended to
squarks and if we take $M_3=r_3 M_2\;\;\; r_3>1$ at the GUT scale
one could make the squarks ``naturally heavy" as we have assumed.
Note also in this respect that had we not taken the squarks,
specifically the stops, sufficiently heavy we would not have had
large enough radiative corrections to the Higgs mass and would
have been in conflict with the LEP2 constraint on the Higgs mass.
Since the limit on the relic density in these scenarios with
$M_1=M_2/10$  constrain essentially the right slepton mass, this
means that one has an almost direct limit on $m_0$.

Putting all this together the parameter space still allowed by
requiring that the relic density be such that $\Omega_\chi h^2
<.3$ and by taking into account all accelerator constraints listed
in section~2 is shown in Fig.~\ref{relic-nounif}. The most
important message is that sleptons must be lighter than about
$140$GeV. The approximate rule of thumb given by
Eq.~\ref{approx-relic} is therefore quite good and explains the
various behaviours of Fig.~\ref{relic-nounif}. Had we imposed a
lower $\Omega_\chi h^2$, $\Omega_\chi h^2<.2$ would have meant
$\mser<125$GeV. Even with the very mild constraint $\Omega_\chi
h^2<.6$ right selectron masses are below $160$GeV. The same figure
also shows the effect of not taking into account the constraint
from the LEP2 neutralino cross sections. As expected the latter
cut on smallest $\mu$ values (and also a bit on smaller $M_2$
values), that not only allow accessible $\neutt,\neutth$ but also
cut on the amount of the higgsino component in $\neuto$ and thus
on the contribution of $\neuto$ to the invisible decay of $h$. We
therefore see that a combination of LEP2 neutralino cross sections
with improved constraints from the relic density are important.

\subsection{Light Sleptons}
We now allow for light sleptons with masses such that
$m_{\tilde{l}}>90$GeV but take into account all cosmological and
accelerator constraints. The masses are calculated according to
Eq.~\ref{m0runing}. Although one has reduced the $\mu-M_2$
parameter space somehow one has also allowed for light sleptons
that indirectly contribute to $h\ra \gamma \gamma$ beside the
light charginos. Right and left charged sleptons of equal masses
contribute almost equally and interfere destructively with the
dominant $W$ loop hence reducing the width $h \ra \gamma \gamma$.
Once again large drops are possible with  reduction factor as
small as $.3$ in both the branching ratio of the Higgs into
photons and $b\bar b$.

\begin{figure*}[htbp]
\begin{center}
\includegraphics[width=16cm,height=15cm]{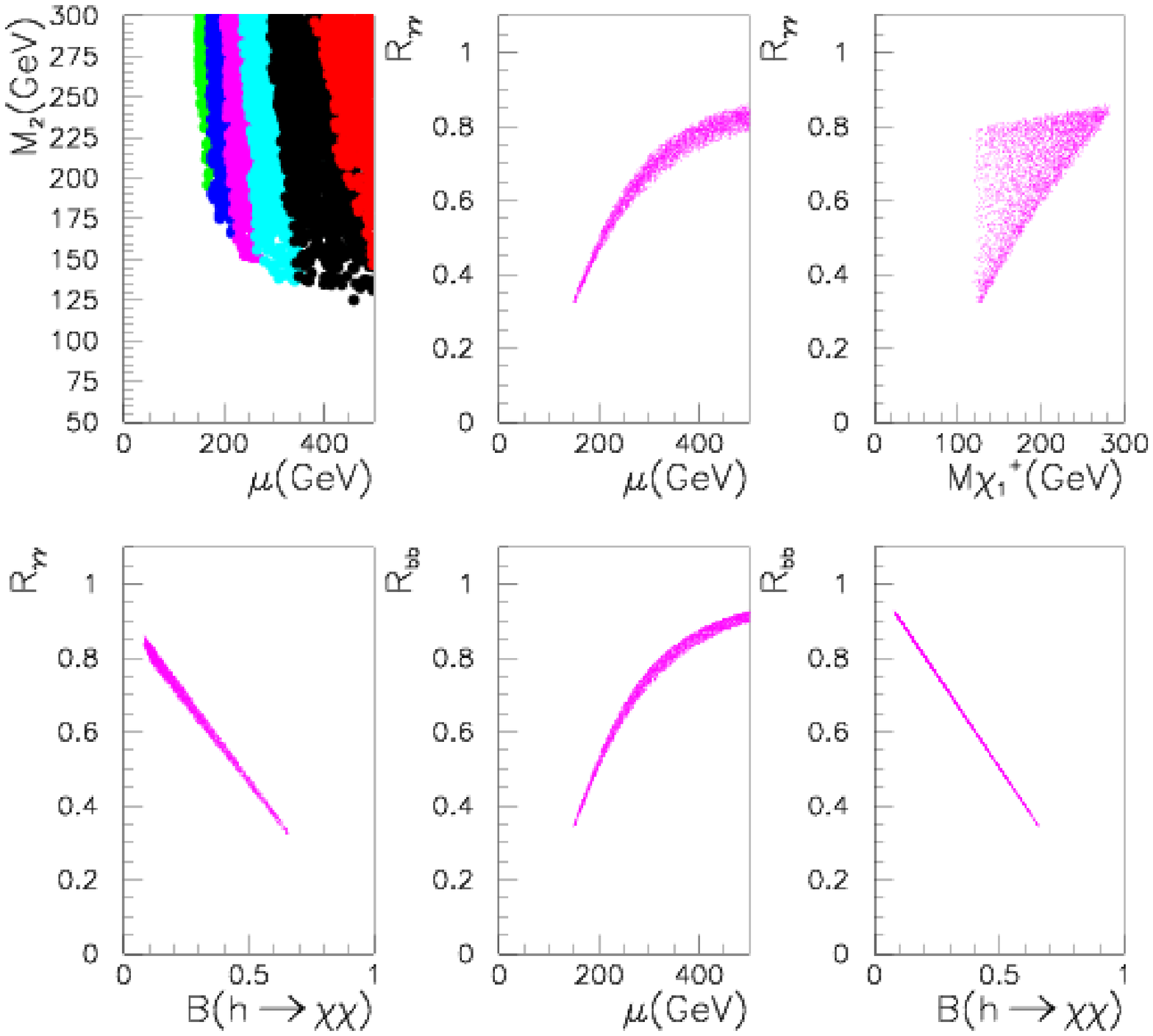}
\caption{\label{nounif-lightslep} {\em Effects of neutralinos from
$M_1=M_2/10$ with $\tgb=5$ and $A_t=0$ with light selectrons. In
all plots the scans are over $M_2=50-300$GeV, $\mu=100-500$GeV.
From left to right and top to bottom \newline a) Density plot for
\rggt  in the allowed $M_2-\mu$ plane. The different shadings
correspond to $ .3 <\rgg <.4$ (left band) to $ .8 <\rgg <.9$
(right band). b) Variation of \rggt$\;$ with $\mu$ c) with the
mass of the chargino $M_{\chi_1^+}$. d) Correlation between \rggt
and the branching into LSP. e) Variation of $R_{b \bar b}$ with
$\mu$. f) $R_{b\bar b}$ {\it vs} the branching of the Higgs into
the LSP neutralino. \/.}}
\end{center}
\end{figure*}

The loop effects of the sleptons are rather marginal compared to
the effect of the opening up of the neutralino channels. They
account for some $10-15\%$ drop as can be seen when $h \ra \neuto
\neuto$ is closed and by comparing with the heavy slepton case.

\subsection{Associated chargino and neutralino production}
In cases where there are very large reductions in the usual $b
\bar b$ and $\gamma \gamma$ signatures of the Higgs,  production
of charginos and neutralino at the LHC is quite
large\footnote{Production of light sleptons, as constrained from
cosmology in these scenarios, is on the other hand quite modest at
the LHC.}.
\begin{figure*}[hbtp]
\begin{center}
\includegraphics[width=10cm,height=10cm]{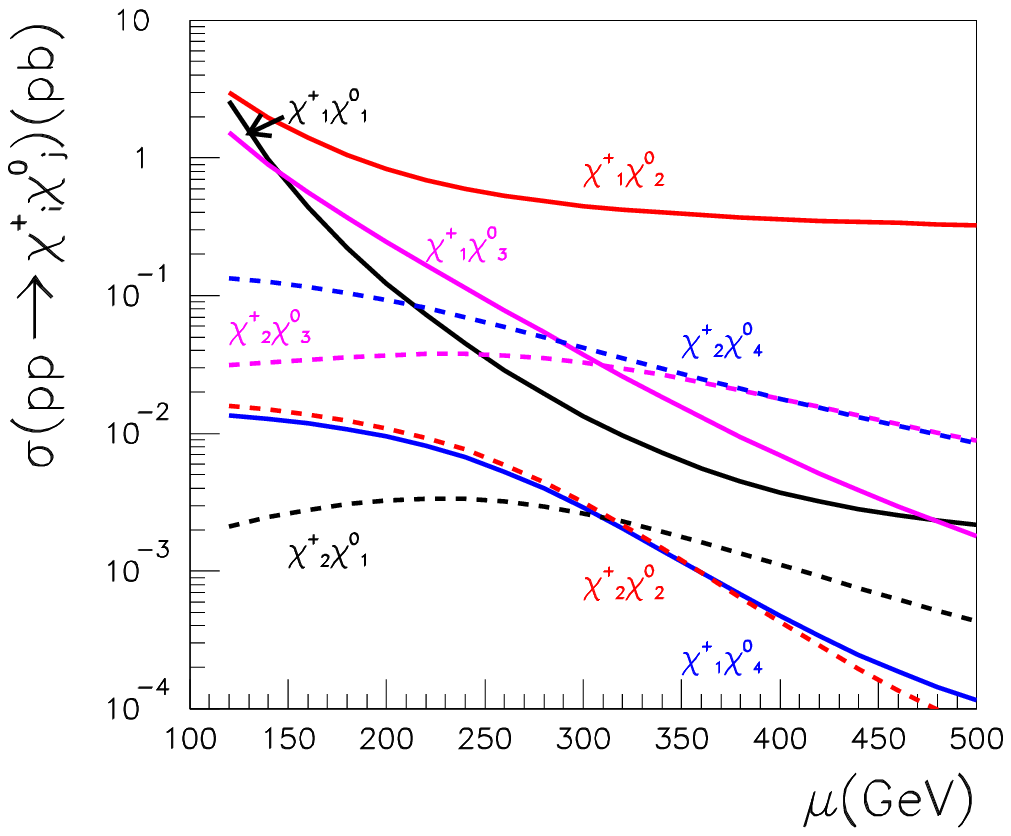}
\includegraphics[width=10cm,height=10cm]{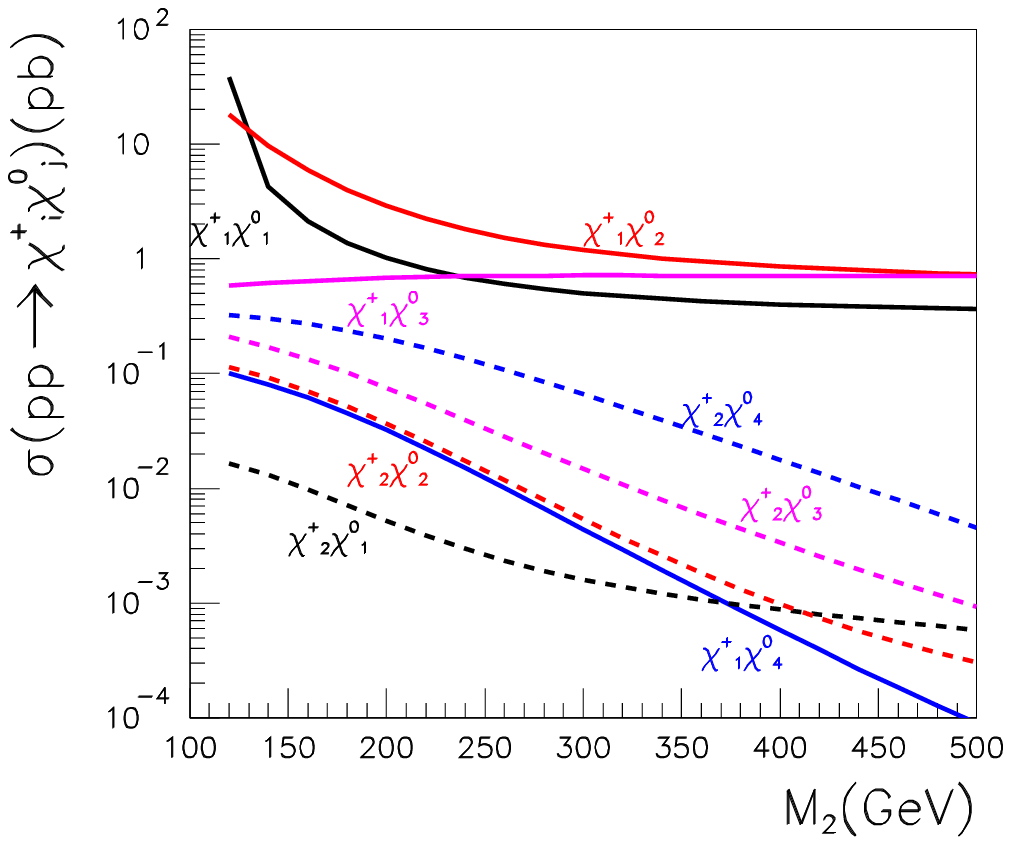}
\caption{\label{chiprod-nounif}{\em Associated Production of
chargino and neutralino at the LHC at LO for $M_2=M_1/10$ a) as a
function of $\mu$ for $M_2=250$GeV. b) as a function of $M_2$ for
$\mu=150$GeV \/.}}
\end{center}

\end{figure*}
Fig.~\ref{chiprod-nounif} shows that, for values of $\mu-M_2$
where $R_{\gamma \gamma}$ is below $.6$, all neutralinos and
charginos can be produced. For instance with $M_2=250$GeV, the
cross section for $\neutf \chi_2^+$ is in excess of $100$fb while
$\neutt \chi_1^+$ is above $1pb$. Therefore early observations of
these events, could probably allow the determination of the
parameters of the higgsino-gauginos sector ``sending an early
warning signal" that indicates difficulty in the detection of the
Higgs.
\begin{figure*}[hbtp]
\begin{center}
\includegraphics[width=15cm,height=10cm]{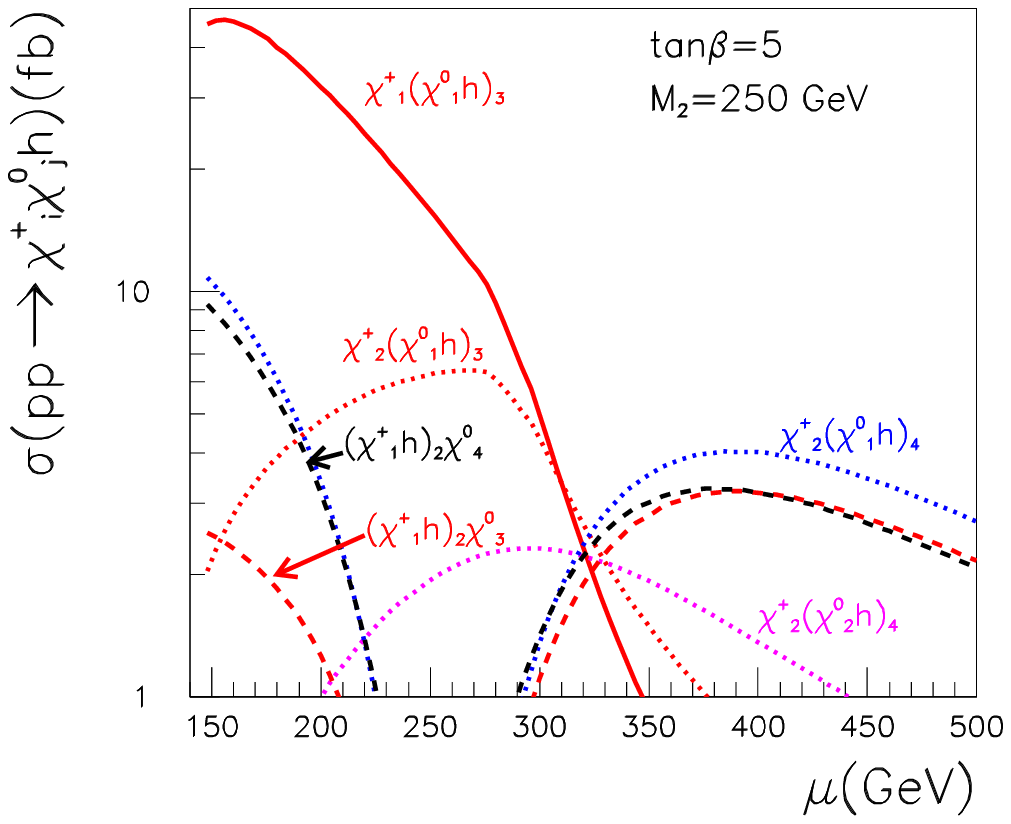}
\caption{\label{folding-nouni}{\em As in
Fig.~\ref{folding-uni-nondeg} but with $M_2=250$GeV and $
M_1=M_2/10$\/.}}
\end{center}
\end{figure*}

If we now look at the (lightest) Higgs that can be produced
through cascade decays in these processes, one sees from
Fig.~\ref{folding-nouni} that, through essentially $\neutth$
decays, associated Higgs cross sections of about $30$fb are
possible. Nonetheless, again, it is in these regions with highest
yield that the Higgs has a large branching ratio into invisible
and would be difficult to track.

\section{Conclusions}
In a model that assumes the usual common gaugino mass at the GUT
scale and where, apart from the charginos and neutralinos, all
other supersymmetric particles are heavy, we have shown that
current LEP limits on charginos imply that there should be no
problem finding the lightest SUSY Higgs at the LHC in the
two-photon mode or even $b \bar b$ in the associated $t\bar t h$
channel. The loop effects of charginos in the two-photon width are
small compared to the theoretical uncertainties, they amount to
less than about $15\%$ and can either increase or decrease the
signal. The LEP data in this scenario mean that the decay of the
Higgs into invisibles is almost closed.  In scenarios ``on the
fringe" with a conspiracy between the sneutrino mass and the
lightest chargino mass,  the Higgs signal can be very much
degraded in both the two-photon and the $b$ final states. This is
because the (invisible) Higgs decay into light neutralinos may
become the main decay mode, suppressing all other signatures. This
also occurs in models that do not assume the GUT inspired gaugino
mass, specifically those where, at the weak scale,  the $U(1)$
gaugino mass is much smaller than the $SU(2)$ gaugino mass.
However we point out that limits from the relic density in these
types of models require rather light right selectron masses. These
in turn contribute quite significantly to the cross section for
neutralino production at LEP2 which then constrains the parameter
space in the gaugino-higgsino sector where the invisible branching
ratio of the Higgs becomes large. Although large reductions in the
usual channels are still possible, the combination of LEP2 data
and cosmology means that observation of the Higgs signal at the
LHC is jeopardised in only a small region of the SUSY parameter
space. Moreover, we show that in these scenarios where the drops
in the Higgs signals are most dramatic, one is assured of having a
quite healthy associated chargino and neutralino cross section at
the LHC. Some of the heavier of these particles  may even trigger
Higgs production through a cascade decay into their lighter
partner. It rests to see whether the Higgs can be seen in this new
production channels, considering that it will predominantly have
an ``invisible" signature.

\vspace*{1cm}

\noi {\bf Acknowledgments}

We would like to thank Sacha Pukhov and Andrei Semenov for help
and advice on the use of {\tt CompHEP}. R.G. acknowledges the
hospitality of LAPTH where this work was initiated. This work is
done under partial financial support of the Indo-French
Collaboration IFCPAR-1701-1 {\em Collider Physics}.


\end{document}